\documentclass[twocolumn,aps,pra,floatfix,tightenlines,superscriptaddress,nofootinbib]{revtex4-1}
\usepackage[T1]{fontenc}
\usepackage{CJKutf8}
\usepackage{color}
\usepackage{verbatim}
\usepackage{mathtools}
\usepackage{amsmath}
\usepackage{graphicx}
\usepackage{esint}
\usepackage[unicode=true,
 bookmarks=false,
 breaklinks=false,pdfborder={0 0 1},backref=false,colorlinks=true]
 {hyperref}
\hypersetup{
 linkcolor=[rgb]{0,0,1},citecolor=[rgb]{0,0,1},urlcolor=[rgb]{0,0,1}}

\makeatletter
\usepackage{amsfonts}
\usepackage{bm}
\usepackage{bbold}
\usepackage{comment}
\usepackage{times}
\usepackage{mathtools}

\usepackage{array}
\newcolumntype{C}[1]{>{\centering\arraybackslash$}p{#1}<{$}}

\newcommand{\1}{\leavevmode{\rm 1\ifmmode\mkern  -4.8mu\else\kern -.3em\fi I}}
\makeatother

\begin{document}
\title{Topological Protection of Coherence in Noisy Open Quantum Systems}
\author{Yu Yao$^{*}$}
\affiliation{Department of Physics and Astronomy, University of Southern California,
Los Angeles, CA 90089-0484}
\author{Henning Schlömer$^{*}$}
\affiliation{Institute for Theoretical Solid State Physics, RWTH Aachen University,
Otto-Blumenthal-Str. 26, D-52056 Aachen, Germany}
\author{Zhengzhi Ma}
\affiliation{Department of Physics and Astronomy, University of Southern California,
Los Angeles, CA 90089-0484}
\author{Lorenzo Campos Venuti}
\affiliation{Department of Physics and Astronomy, University of Southern California,
Los Angeles, CA 90089-0484}
\author{Stephan Haas}
\affiliation{Department of Physics and Astronomy, University of Southern California,
Los Angeles, CA 90089-0484}
\date{\today}
\begin{abstract}
We consider topological protection mechanisms in dissipative quantum
systems in the presence of quenched disorder, with the intent  to prolong coherence times of qubits. The
physical setting is a network of qubits and dissipative cavities whose
coupling parameters are tunable, such that topological edge states
can be stabilized. The evolution of a fiducial qubit is entirely determined by a non-Hermitian Hamiltonian which thus emerges from a bona-fide physical process. It is shown how even in the presence of disorder
winding numbers can be defined and evaluated in real space, as long
as certain symmetries are preserved. Hence we can construct the topological
phase diagrams of noisy open quantum models, such as the non-Hermitian
disordered Su-Schrieffer-Heeger dimer model and a trimer model that
includes longer-range couplings. In the presence of competing disorder
parameters, interesting re-entrance phenomena of topologically non-trivial
sectors are observed. This means that in certain parameter regions, increasing disorder drastically increases the coherence time of the fiducial qubit. 
%It is argued that based on these open quantum
%networks, topologically protected gates can be achieved via proper
%braiding. 
\end{abstract}
\maketitle
\global\long\def\thefootnote{*}%
\footnotetext{These authors contributed equally to this work}

\section{Introduction}
Due to their characteristic protection against environmental noise, %edge modes in 
topological phases of matter \citep{Fu2007,Fu2007_2, Kobayashi2013} are considered to be promising candidates for the realization of noise-resilient quantum computers \citep{Sankar2008, Sau2010, Sarma2015, Stern2013, Alicea2011, Freedman2006, Freedman2003, Akhmerov2010}. 
Furthermore, it was shown in \cite{Venuti2017} that the presence of topological edge states can preserve quantum mechanical features, e.g. coherence of a fiducial qubit, in the presence of dissipation (see also \citep{man_cavity-based_2015, campos_venuti_long-distance_2006, campos_venuti_qubit_2007} for other works in a similar spirit).
In that work, %\cite{Venuti2017}, 
dissipative one-dimensional (1D) quantum optical qubit-cavity architectures were analyzed, where  effective non-Hermitian Hamiltonians of the form of a tight-binding chain with diagonal complex entries were derived. The time evolution of the boundary-qubit coherence, driven by such non-Hermitian Hamiltonians, was extensively studied for choices of hopping parameters that admit symmetry-protected topological states localized at the edges of the system. It was found that (quasi-)dark modes, i.e., boundary states with  exponentially small (in system size) imaginary parts, protect the edge qubits from decoherence effects via photon leakage through cavities. 

Moreover, disordered as well as non-Hermitian generalizations of 1D topological insulators, such as the Su-Schrieffer-Heeger (SSH) model \citep{Su1980, Heeger1988}, have been studied theoretically \citep{Mondragon2014,Luo2019}, where the real space winding number was analyzed for different disorder strengths on the hopping parameters. 

Here, we build on the work presented in \cite{Venuti2017},  addressing the role of quenched disorder in the qubit-cavity arrays. We fully characterize the disordered, non-Hermitian system's topology by computing the winding number in real space in the parameter space spanned by the coupling amplitude and the disorder strength. From this characterization, accurate predictions for the behavior of a fiducial qubit's coherence can be made for long times, therefore expanding the discussion of the quantum optical systems to a broader physical context, considering both quenched disorder and dissipation.

The remainder of this paper is organized as follows. In Sec.~\ref{sec:setup}, we  derive the effective non-Hermitian Hamiltonian describing qubit-cavity arrays using the Lindblad formalism. In Sec.~\ref{sec:real_space_method}, the topological characterization of dissipative, disordered systems is illustrated, which is then applied to non-Hermitian dimer and trimer models in Sec~\ref{sec:main}. Special focus is put on the coherence of the qubit located at the boundary, whose faith can be accurately predicted from the phase diagrams. We then briefly discuss possible applications in quantum computation via dark-state braiding in Sec.~\ref{sec:quantum_computation} and conclude in Sec.~\ref{sec:conclusion}.

\section{The Setup}

\label{sec:setup} 
\begin{figure}
\begin{centering}
\includegraphics[width=0.48\textwidth]{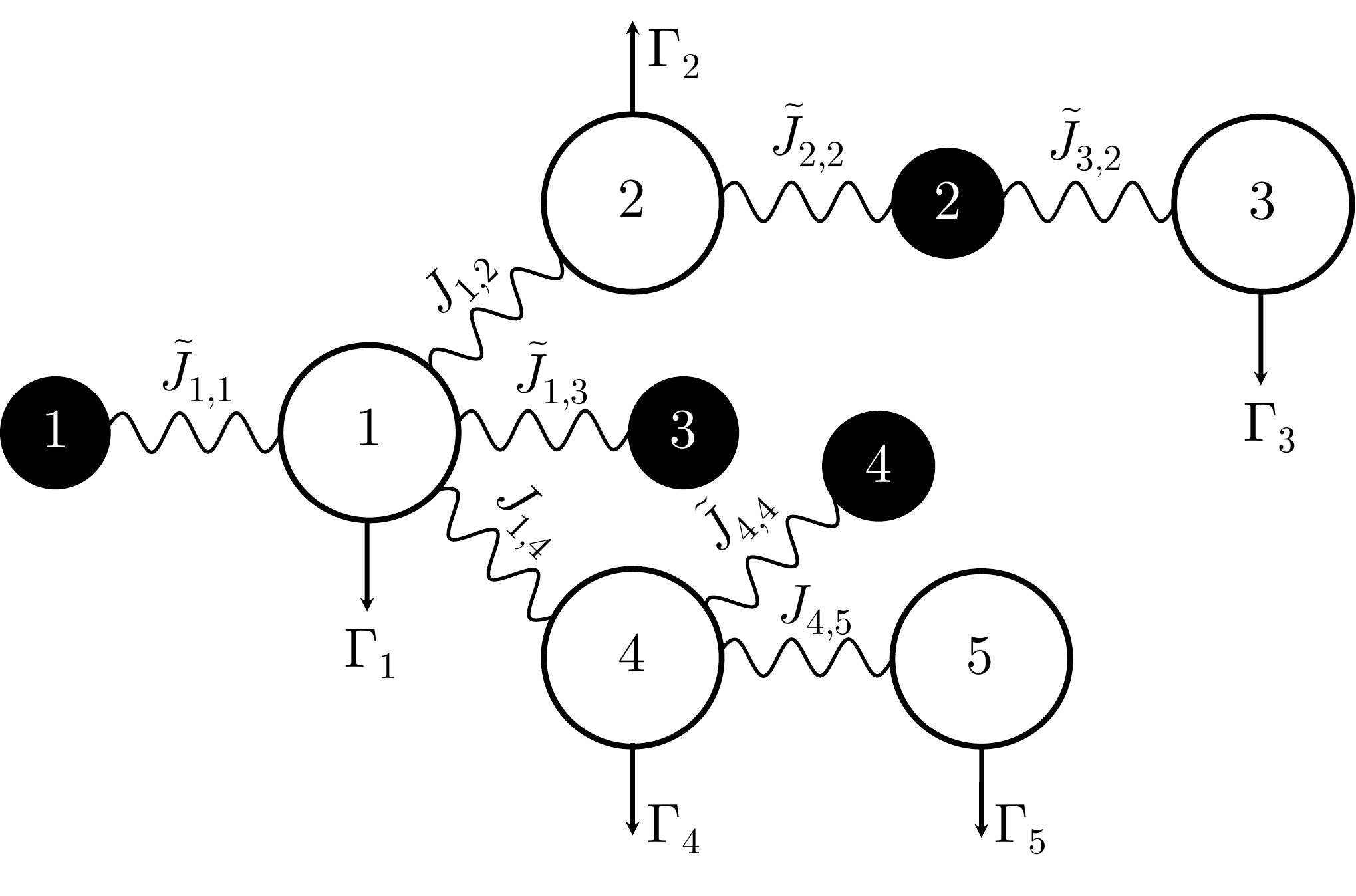} \caption{Example of a network of qubits (filled circles) interacting with lossy
cavities (hollow circles). Wavy lines indicate coherent hopping matrix
elements $\tilde{J}_{i,j}$, connecting cavity $i$ with qubit $j$,
and $J_{i,j}$, connecting cavity $i$ with cavity $j$. Arrows indicate
incoherent decay $\Gamma_{i}$ in cavity $i$. \label{fig:network} }
\par\end{centering}
\end{figure}

We consider a network consisting of qubits coupled
to dissipative cavities in a Jaynes-Cumming fashion. Specifically,
we study networks of $M$ qubits and $K$ dissipative cavities, as
illustrated in Fig. \ref{fig:network} for $M=4$ and $K=5$. The
Hamiltonian of the system has the following form 
\begin{eqnarray}
H & = & \sum_{l,m=1}^{K}J_{l,m}(a_{l}^{\dagger}a_{m}+\text{h.c.})\nonumber \\
 & + & \sum_{i=1}^{M}\sum_{l=1}^{K}\tilde{J}_{l,i}(a_{l}^{\dagger}\sigma_{i}^{-}+\text{h.c.}),\label{hamiltonian}
\end{eqnarray}
where $a_{l}^{\dagger}$ and $a_{l}$ are the bosonic creation and
annihilation operators for cavity mode $l$, and $\sigma_{i}^{\pm}$
are the ladder operators for qubit $i$. We consider a Lindblad master
equation $\dot{\rho}=\mathcal{L}[\rho]$, where $\mathcal{L}=\mathcal{K}+\mathcal{D}$,
the coherent part is $\mathcal{K}=-i\left[H,\bullet\right]$, whereas
the dissipative part is 
\begin{equation}
\mathcal{D}[\rho]=\sum_{l=1}^{K}\Gamma_{l}\Big[2a_{l}\rho a_{l}^{\dagger}-\{a_{l}^{\dagger}a_{l},\rho\}\Big].\label{lindbladian}
\end{equation}
Such Lindbladian description is accurate at sufficiently low temperature
in particular in circuit QED experiments. 

The Lindbladian can also be written as $\mathcal{L}=\mathcal{K}'+\mathcal{D}'$,
where $\mathcal{K}'(\rho)=-i(H'\rho-\rho{H'}^{\dagger})$ defines
the non-Hermitian Hamiltonian $H'$: 
\begin{equation}
H'=H-i\sum_{l=1}^{K}\Gamma_{l}a_{l}^{\dagger}a_{l},
\end{equation}
and 
\begin{equation}
\mathcal{D}'(\rho)=\sum_{l=1}^{K}2\Gamma_{l}a_{l}\rho a_{l}^{\dagger}.
\end{equation}

Consider the Fock space of the system $\mathcal{F}=\oplus_{n=0}^{\infty}\mathcal{H}_{n}$
where $\mathcal{H}_{n}$ is the Hilbert space of $n$ particles (at
this level the distinction between spins and bosons is unimportant).
In the space of operators on $\mathcal{F}$ we define the Hilbert-Schmidt
scalar product $\langle\!\langle x|y\rangle\!\rangle=\operatorname{Tr}(x^{\dagger}y)$.
Using the isomorphism $\mathcal{B}_{HS}(\mathcal{F})\simeq\mathcal{F}\otimes\mathcal{F}^{*}$
($\mathcal{B}_{HS}(\mathcal{F})$ is the space of bounded Hilbert-Schmidt
operators on $\mathcal{F}$), the space of operators $\mathcal{B}_{HS}(\mathcal{F})$
can be identified with 
\begin{equation}
\mathcal{B}_{HS}(\mathcal{F})\simeq\bigoplus_{i,j=0}^{\infty}\mathcal{H}_{i}\otimes\mathcal{H}_{j}^{*}.
\end{equation}
In simpler terms, $\mathcal{B}_{HS}(\mathcal{F})$ has a block structure
with two labels $(i,j)$ each label being a particle number. The non-Hermitian
Hamiltonian $H'$ preserves the number of particles and correspondigly
$\mathcal{K}'$ is block-diagonal in $(i,j)$. Instead, $\mathcal{D}'$
connects the sector $(i,j)$ with the sector $(i-1,j-1)$, i.e.~it
decreases the number of particles by one.

In this paper we will be mostly interested in the coherence of a fiducial
qubit, that, without loss of generality we place at site 1. In the
standard basis, the coherence of a qubit in state $\rho$, can be
defined as $\mathcal{C}=2\vert\rho_{\downarrow,\uparrow}\vert$ \citep{Baumgratz2014}.
We initialize the system such that all cavities are empty and qubits
are in the lowest state ($|\downarrow\rangle$), while on the fiducial
qubit the state is $|\psi\rangle=\alpha|\downarrow\rangle+\beta|\uparrow\rangle$.
We further fix $|\alpha\beta|=1/2$ which means that at the beginning
the coherence assumes its maximal value one. We are interested in
the evolution of the coherence as a function of time.

Let $|0\rangle$ be the vacuum state with no excitation on any qubit
or cavity, while $|j\rangle$ denotes a single excitation on the $j$th
site, describing either an excited qubit or a cavity hosting a photon.
We use the notation $|j\rangle\!\rangle\leftrightarrow|0\rangle\langle j|$.
It can be shown \cite{Venuti2017} that the evolution of the coherence at later
time is given by 
\begin{eqnarray}
\mathcal{C}(t)=2|\rho_{\downarrow,\uparrow}(t)| & = & 2|\langle0|\rho(t)|1\rangle|=2|\operatorname{Tr}|1\rangle\langle0|\rho(t)|\nonumber \\
 & = & 2|\langle\!\langle1|\rho(t)\rangle\!\rangle|=2|\langle\!\langle1|e^{t\mathcal{L}}|\rho(0)\rangle\!\rangle|.\label{coh}
\end{eqnarray}

Because of the block structure of the Lindbladian, $\langle\!\langle1|e^{t\mathcal{L}}|\rho(0)\rangle\!\rangle=\langle\!\langle1|e^{t\tilde{\mathcal{L}}}|\tilde{\rho}(0)\rangle\!\rangle$,
where $\tilde{X}$ is the operator $X$ restricted to the linear space
$\mathcal{V}_{0,1}=\mathrm{Span}(|0\rangle\langle j|),j=1,2,\ldots,N$.
In particular, $\tilde{\rho}(0)=\overline{\alpha}\beta|0\rangle\langle1|=(1/2)|0\rangle\langle1|$.
For what regard the restriction of the Lindbladian we have $\tilde{\mathcal{L}}=\tilde{\mathcal{K}'}$.
Moreover, in $\mathcal{V}_{0,1}$,
\begin{align}
{\mathcal{K}'}_{l,m} & =\operatorname{Tr}\left(|l\rangle\langle0|{\mathcal{K}'}(|0\rangle\langle m|)\right)\\
 & =i\langle m|{H'}^{\dagger}|l\rangle\\
 & =i\overline{\langle l|H'|m\rangle}.
\end{align}

In other words, the evolution of the coherence of the fiducial qubit
is entirely determined by the non-Hermitian Hamiltonian $-\overline{H'}$
in the one-particle sector. Calling $\mathsf{H}:=-\left.\overline{H'}\right\vert _{\mathrm{one\ particle}}$
we have finally

\begin{equation}
\mathcal{C}(t)=\left\vert \langle\!\langle1|e^{-it\mathsf{H}}|1\rangle\!\rangle\right\vert .\label{eq:coherence_final}
\end{equation}

We would like to stress that, in this setting, a non-Hermitian Hamiltonian
emerges from a genuine \emph{bona-fide} quantum evolution whereas
in most current proposals non-Hermitian Hamiltonians are simulated
in classical dissipative wave-guides via the analogy between Helmoltz
and Schrödinger equation (see e.g.~\citep{rudner_topological_2009}).

In order to prolong the coherence Eq.~\eqref{eq:coherence_final},
one seeks a (non-Hermitian) Hamiltonian $\mathsf{H}$ that admits i)
long-lived states, i.e.~eigenstates of $\mathsf{H}$ with small (negative)
imaginary part; and ii) that also have large amplitude on the site
$|1\rangle\!\rangle$ (conventionally placed at the beginning of the
chain). 

Interestingly, both these requirement are satisfied to a large degree
in one dimensional topological systems which admit edge states with
the required properties in the non-trivial phase. The classification
of such dissipative, non-Hermitian, topological chains has been done
in \citep{Levitov2016} and utilized to prolong quantum coherence
for the first time in \citep{Venuti2017}. Here we extend the analysis
of \citep{Venuti2017} to disordered systems where translational invariance
is broken. The phase diagrams of topological dissipative chains will
tell us which parameter regions and models can be used to prolong
the quantum coherence of the fiducial qubit.

\section{Topological invariant of dissipative systems in real space}

\label{sec:real_space_method}

We now briefly recall the topological classification of non-Hermitian
quantum systems provided by Rudner \textit{et al.} in Ref.~\citep{Levitov2016}.
For non-Hermitian quantum systems hosting dissipative sites, the
topological invariant can be defined as the winding number around
the dark-state manifold in the Hamiltonian parameter space. A non-trivial
phase in dissipative systems corresponds to long-lived edge modes
with infinite or exponential large lifetimes.

In previous work \citep{Venuti2017}, the topological classification
of non-Hermitian models was formulated within the framework of Bloch
theory, which we briefly outline here for comparison with the real-space
approach to be introduced. Consider a one-dimensional periodic non-Hermitian
chain with $n$ sites per unit cell. In the thermodynamic limit, the
Hamiltonian is given by $\mathsf{H}=\oint dk/(2\pi)\sum_{\alpha,\beta=1}^{n}H_{\alpha,\beta}(k)|k,\alpha\rangle\langle k,\beta|$.
We shall only focus on the cases with one leaky site per unit cell,
as the topological characterization is trivial in all other cases
if no additional constraints are imposed \citep{Levitov2016}.
The Bloch Hamiltonian of any such system is an $n\times n$ matrix,
which can be written as 
\begin{equation}
H(k)=\left(\begin{array}{cc}
h(k) & v_{k}\\
v_{k}^{\dagger} & \Delta(k)-i\Gamma
\end{array}\right),
\end{equation}
where $h(k)$ is an $(n-1)\times(n-1)$ Hermitian matrix, $v_{k}$
is a $(n-1)$-dimensional vector and $\Delta(k)-i\Gamma$ is a complex
number. The Hamiltonian can be further decomposed in the following
manner 
\begin{equation}
H(k)=\left(\begin{array}{cc}
U(k) & 0\\
0 & 1
\end{array}\right)\left(\begin{array}{cc}
\tilde{h}(k) & \tilde{v}_{k}\\
\tilde{v}_{k}^{\dagger} & \Delta(k)-i\Gamma
\end{array}\right)\left(\begin{array}{cc}
U(k)^{\dagger} & 0\\
0 & 1
\end{array}\right),\label{eqn:Hk}
\end{equation}
where $U(k)$ is a $(n-1)\times(n-1)$ unitary matrix whose columns
are the eigenvectors of $h(k)$, and $\tilde{h}(k)$ is the $(n-1)\times(n-1)$
diagonal matrix of the corresponding eigenvalues. The phases of the
eigenvectors are fixed by making all entries of the $(n-1)$-dimensional
vector $\tilde{v}_{k}$ real and positive. Any $U(k)$ satisfying
the above criteria can be chosen without affecting the final result.
Since $U(k)$ is the only component parametrizing the Hamiltonian
that can lead to non-trivial topology \citep{Levitov2016}, the winding
number of $H(k)$ reduces to the one of $U(k)$, which is given by
\begin{equation}
W=\oint\frac{dk}{2\pi i}\partial_{k}\ln\operatorname{det}\{U(k)\}.\label{eqn:Wk}
\end{equation}
We now construct a real-space representation of the winding number
that remains well defined when translation invariance is destroyed
by e.g.~the presence of disorder. Consider a chain with $n$ sites
in each cell and $M$ number of unit cells. For what we said previously,
we consider only one leaky site per unit cell, which, without loss
of generality, we place at the final site of the cell. % there exists one decaying sublattice which we denote by $\zeta$. The real-space Hamiltonian can be rearranged in the order of sublattices, %which can be represented as

The one-particle (non-Hermitian) Hamiltonian can be written as

\begin{equation}
\mathsf{H}=\sum_{i,j=1}^{M}\sum_{\alpha,\beta=1}^{n}H_{\alpha,\beta}^{i,j}|i,\alpha\rangle\langle j,\beta|\label{eq:H_one}
\end{equation}

%\begin{eqnarray}%H=\left(\begin{array}{ccccc}%\epsilon_{A} & H_{AB} & H_{AC} & \dots & H_{A\zeta} \\%H_{BA} & \epsilon_{B} & H_{BC} & \dots & H_{B\zeta} \\%H_{CA} & H_{CB} & \epsilon_{C} & \dots & H_{C\zeta} \\%\vdots & \vdots & \vdots & \ddots \\%H_{\zeta A} & H_{\zeta B} & H_{\zeta C} & & \epsilon_{\zeta} - i \Gamma%\end{array}\right).%\end{eqnarray}

Generally one thinks of the chain as being made of $M$ cells with $n$
sites each, but one may as well think of $n$ sections with $M$ sites
each. In other words, we rearrange Eq.~\eqref{eq:H_one} according
to the following block structure
\begin{eqnarray}
\mathsf{H}=\left(\begin{array}{ccccc}
H_{1,1} & H_{1,2} & H_{1,3} & \dots & H_{1,n}\\
H_{2,1} & H_{2,2} & H_{2,3} & \dots & H_{2,n}\\
\vdots & \vdots & \vdots &  & \vdots\\
H_{n,1} & H_{n,2} & H_{n,3} & \dots & H_{n,n}
\end{array}\right),
\end{eqnarray}
where each $H_{\alpha,\beta}$ is a $M\times M$ matrix. The matrices
$H_{\alpha,\alpha}$ $\alpha=1,\ldots,(n-1)$ are diagonal with chemical
potentials on the diagonal. Since we put the leaky site at position
$\alpha=n$, the matrix $H_{n,n}=\epsilon_{n}-i\Gamma\leavevmode{\rm 1\ifmmode\mkern-4.8mu \else\kern-0.3em \fi I}$, 
where $\epsilon_{n}$ is a diagonal matrix of chemical potentials
and for simplicity we set the leakage to have value $\Gamma$ on each
site.

%Here, $\epsilon_{\alpha}$ depicts the on-site chemical potential of sublattice $\alpha$, and $H_{\alpha\beta}$ describes the hopping between sublattices $\alpha$ and $\beta$. 

Recalling the approach used in $k$-space, we first write the real-space
Hamiltonian as 
\begin{align}
H & =\left(\begin{array}{cc}
\Lambda & V\\
V^{\dagger} & \epsilon_{n}-i\Gamma\leavevmode{\rm 1\ifmmode\mkern-4.8mu \else\kern-0.3em \fi I}
\end{array}\right)\nonumber \\
 & =\left(\begin{array}{cc}
U & 0\\
0 & \leavevmode{\rm 1\ifmmode\mkern-4.8mu \else\kern-0.3em \fi I}
\end{array}\right)\left(\begin{array}{cc}
\tilde{\Lambda} & \tilde{V}\\
\tilde{V}^{\dagger} & \epsilon_{n}-i\Gamma\leavevmode{\rm 1\ifmmode\mkern-4.8mu \else\kern-0.3em \fi I}
\end{array}\right)\left(\begin{array}{cc}
U^{\dagger} & 0\\
0 & \leavevmode{\rm 1\ifmmode\mkern-4.8mu \else\kern-0.3em \fi I}
\end{array}\right).\label{eqn:Hr}
\end{align}
%where we divide $H$ into four sectors and separate decaying and non-decaying sublattices. 
$\Lambda$ is a $(n-1)M\times(n-1)M$ Hermitian matrix, while $V$
is a $(n-1)M\times M$ real matrix describing the hopping between
decaying and non-decaying sites. $\tilde{\Lambda}$ is a $(n-1)L\times(n-1)L$
diagonal matrix with real eigenvalues of $\Lambda$, and $U$ is a
$(n-1)L\times(n-1)L$ unitary matrix that diagonalizes $\Lambda$.
The degrees of freedom for the choice of $U$ are fixed by making
each $L\times L$ submatrix in $\tilde{V}$ positive-definite, analogous
to the procedure in reciprocal space. %Note that in real space the statement still holds that the only component in the Hamiltonian that can support any non-trivial topology is $U$, %reducing the winding number of $H$ to the one of $U$.

With these preparations, the winding number of the unitary matrix
$U$ in real space can be evaluated with the prescription of \citep{Kitaev2006}
and further elaborations of Refs.~\citep{Mondragon2014,Wang2019,Luo2019}.
In particular, $\int_{0}^{2\pi}(dk/2\pi)\times\operatorname{tr}\{\}$
and $\partial_{k}$ become trace per volume and the commutator $-i[X,]$
($X$ being the position operator), respectively. Note that $X$ is
the $M$-sized cell position operator, i.e.~$X=\operatorname{diag}(1,2,\ldots,M,1,2,\dots,M,\ldots,M-1,M)$.

Thus, Eq.~(\ref{eqn:Wk}) in real space can be written as 
\begin{equation}
W=\frac{1}{L^{\prime}}\operatorname{tr}^{\prime}(U^{\dagger}[X,U]).\label{eqn:Wr}
\end{equation}
Here, $\operatorname{tr}^{\prime}$ stands for trace with truncation. Specifically, 
we take the trace over the middle interval of length $M^{\prime}$
and leave out $\ell$ sites on each side (total length $M=M^{\prime}+2\ell$).
%As already stated, the chain is divided into $n$ cells of length $M$. Inceach cell we take the trace over the middle interval of length $M^{\prime}$ and leave out $\ell$ sites on each side (total length $M=M^{\prime}+2\ell$).
With Eq.~(\ref{eqn:Wr}), we can explore topological phases in
presence of dissipation and disorder.
%Notably, the classification of non-Hermitian systems can be extended to systems without translational symmetry. %The entire chain is split into  $(n-1) M$ "normal sites" sites and $M$ decaying sites. According to the dimension counting argument raised in %\cite{Levitov2016}, if there exist more than one decaying site per unit cell, the system can only admit non-trivial topological phases when %additional constraints are imposed, i.e., the "strong bipartite" constraint discussed in Ref.~\cite{Levitov2016}.
Note that in the model that we will consider, the matrix $\Lambda$ is not noisy. In general, the model supports a non-trivial topological phase as long as a certain (chiral) symmetry is preserved. Disorder on the elements of $\Lambda$ destroys the symmetry and consequently the system becomes topologically trivial.
%only if no disorder is inflicted on the elements of the matrix $\Lambda$ in Eq.~\ref{eqn:Hr}. 
%For example, disorder acting on the chemical potential of the non-dissipative sites would invalidate the topological characteristics and force the system to be topologically trivial. This is, in fact, supported by our numerical investigations. Due to the additional constraint imposed on the set of allowed Hamiltonians, a non-trivial classification can still emerge, leading to the existence of long-lived edge modes. Such states live solely on the non-decaying sublattice, corresponding to dark states in the non-Hermitian quantum system. Below, we will see that these dark edge modes maintain robustness against disorder, as they do in Hermitian topological systems. \textbf{what is this?}

\section{Disordered non-Hermitian Systems}

\label{sec:main}

We now apply the real space formalism to investigate topological
features in two explicit network geometries, namely the disordered
non-Hermitian SSH dimer model and a disordered non-Hermitian trimer
model.

\subsection{Disordered non-Hermitian SSH Dimer Model}
\label{sec:ssh} 
\begin{figure}
\begin{centering}
\includegraphics[width=0.48\textwidth]{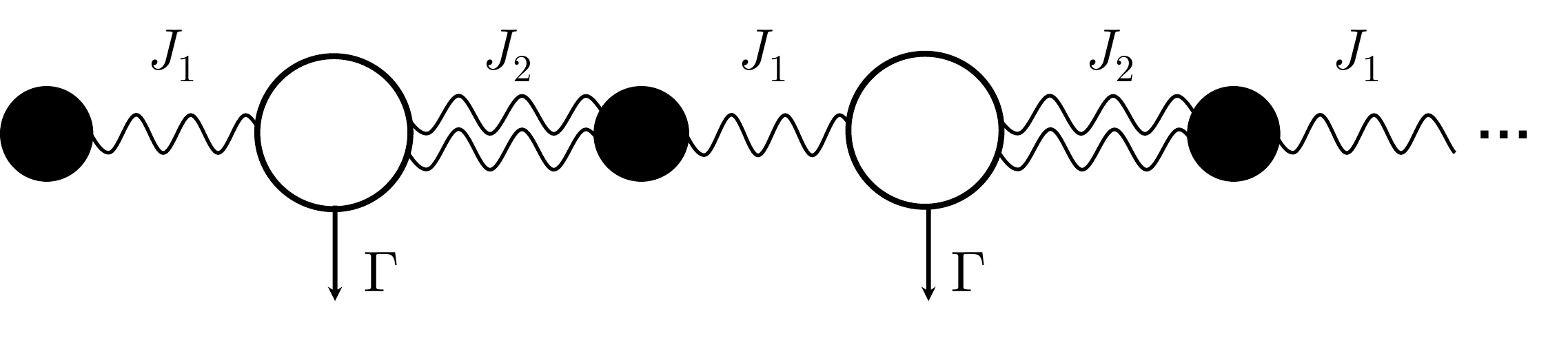} \caption{Non-Hermitian Su-Schrieffer-Heeger (SSH) model with open boundary
conditions. The qubit-cavity and cavity-qubit couplings are given
by $J_{1}$ and $J_{2}$ respectively. Off-diagonal disorder is controlled
via a uniform distribution function from which the hopping parameters
are drawn. \label{fig:ssh}}
\par\end{centering}
\end{figure}

This model describes an open quantum system of coupled qubits and
optical cavities which are arranged in an alternating manner, as shown
in Fig.~\ref{fig:ssh}. In the super-one-particle sector, the corresponding
restricted Hamiltonian $\mathsf{H}$ in the presence of disorder is
given by
\begin{eqnarray}
\mathsf{H} & = & \sum_{j=1}^{M}\epsilon_{A,j}|j,A\rangle\langle j,A|+(\epsilon_{B,j}-i\Gamma)|j,B\rangle\langle j,B|\nonumber \\
 & + & \sum_{j=1}^{M}(J_{1,j}|j,B\rangle\langle j,A|+\mathrm{h.c.})\nonumber \\
& + & \sum_{j=1}^{M-1}(J_{2,j}|j+1,A\rangle\langle j,B|+\mathrm{h.c.}).\label{2siteH}
\end{eqnarray}
%\begin{equation*}
%\mathsf{H}=\left(\begin{array}{ccccc}
%\epsilon_{A,1} & J_{1,1} & 0 & 0 & 0\\
%J_{1,1} & \epsilon_{B,1}-i\Gamma & J_{2,1} & 0 & 0\\
%0 & J_{2,1} & \epsilon_{A,2} & J_{1,2} & 0\\
%0 & 0 & J_{1,2} & \epsilon_{B,2}-i\Gamma & \ddots\\
%0 & 0 & 0 & \ddots & \ddots
%\end{array}\right).\label{eq:H_SSH}
%\end{equation*}
% $J_{1,j}$ describes the hopping from qubit to cavity in the $j$-th unit cell, whereas $J_{2,j}$ is the hopping from cavity $j$ to the qubit at site $(j+1)$. $\epsilon_{A/B,j}$ describe the on-site potentials on each site, and $\Gamma$ is the dissipation rate of each cavity. In the clean limit, with vanishing on-site potentials, $J_{1,j}=J_1$ and $J_{2,j}=J_2$ for all unit cells.
Due to the chiral symmetry and the pseudo-anti-hermiticity of the
non-dissipative and dissipative model, respectively \citep{Venuti2017,Lieu2018},
the topological states are expected to be robust against the chiral
symmetry preserving off-diagonal disorder, i.e., noise in the hopping
parameters. In contrast, disorder in the on-site potentials breaks
the symmetries and is thus expected to quickly diminish topological
features. %This is further explained by the topological characterization of topology in dissipative quantum systems,
Indeed, diagonal disorder leads to a unit cell as large as the system, thus
having more than one dissipative site per unit cell and hence preventing
the existence of topological dark states according to the argument
in \citep{Levitov2016}. We therefore restrict the randomness to act
on the hopping parameters, i.e., $J_{1,j}\equiv J_{1}+\mu_{1}\omega_{1,j}$
and $J_{2,j}\equiv J_{2}+\mu_{2}\omega_{2,j}$, where $\omega_{\alpha,j}$
are independent random variables with uniform distribution in the
range $[-1,+1]$. %Thus $\mu_{1}$ and $\mu_{2}$ are parameters describing the strength of disorder.

\begin{figure}
\centering \includegraphics[width=0.45\textwidth]{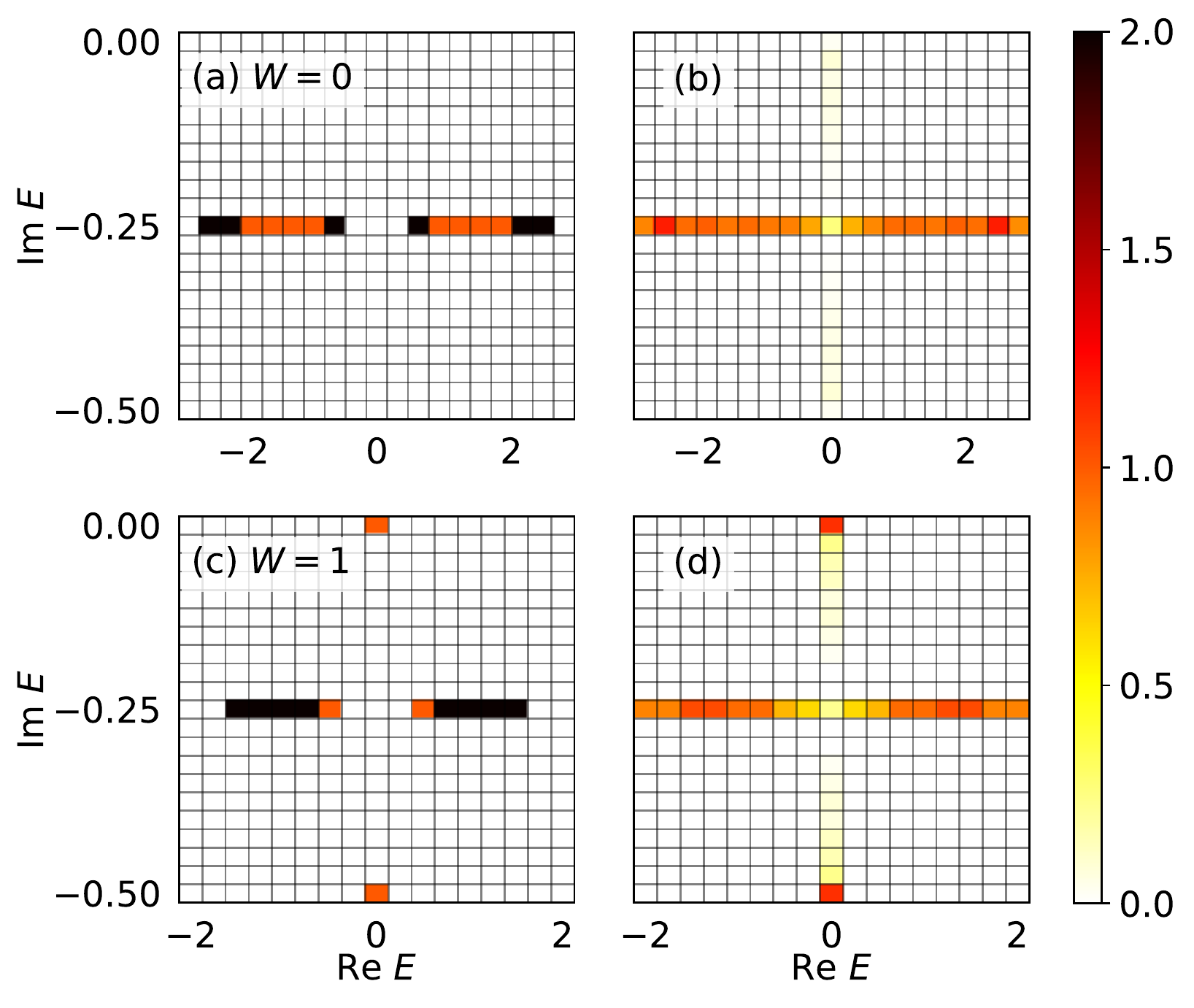}
\caption{Complex density of states of the restricted Hamiltonian $\mathsf{H}$.
(a)\&(b) Topologically trivial regime for the clean and disordered
($\mu=1$) case, respectively. (c)\&(d) Topologically non-trivial
phase for clean and disordered ($\mu=1$) systems, respectively. Results
are averaged over 1000 diagonalizations. Here, $N=20$, $\Gamma=0.5$,
$J_{2}=1$ and $J_{1}=1.5$ ($J_{1}=0.5$) for the topologically trivial
(non-trivial) configurations.}
\label{fig:dos_offdiag} 
\end{figure}

The effect of off-diagonal disorder on the spectrum of the restricted
Hamiltonian $\mathsf{H}$ is illustrated in Figure \ref{fig:dos_offdiag},
where the density of states %for both topological phases considering clean and disordered systems is
is plotted in the complex plane. In the topologically trivial regime of the clean system,
Fig.~\ref{fig:dos_offdiag}~(a), all eigenvalues have imaginary part
$-\Gamma/2$. When disorder is introduced, they mainly wash out on
axis $\mathrm{Im}(E)=-\Gamma/2$, as seen in Fig. \ref{fig:dos_offdiag}(b)
. There is, however, a notable non-vanishing density of states emerging
in the vicinity of $\operatorname{Re}(E)=0$. In the topologically
non-trivial regime, Figs.~\ref{fig:dos_offdiag}~(c)\&(d), a dark
state with corresponding $\operatorname{Im}(E)=0$ can be found. Its
topological protection against off-diagonal disorder manifests itself
in its eigenvalue being left almost unchanged when disorder disturbs
the system, while the bulk states featuring eigenvalues with imaginary
part $-\Gamma/2$ blur out. The protected dark state corresponds to
an edge state having support only on the non-dissipative sites, thus
not decaying through the cavities. Another state emerging in the non-trivial
phase lives, on the contrary, only on the dissipative sites, with
eigenvalue satisfying $\mathrm{Im}(E)=-\Gamma$, as also seen in Fig.~
\ref{fig:dos_offdiag}~(c). The mentioned destructive character of
on-site potential disorder is discussed in the Appendix, Sec.~\ref{sec:diagonal_disorder},
where the density of states for diagonal disorder is analyzed, see
Figs~\ref{fig:diag_disorder}~(a)-(d).

We now turn to the computation of the winding number. In absence of
disorder we can go to reciprocal space and realize that the unitary
$U(k)$ in Eq.~\eqref{eqn:Hk} is simply given by the phase of $J_{1}+J_{2}e^{-ik}$.
The winding number of the dissipative system is thus the same as the
winding number of the closed, Hermitian SSH-chain, resulting in 
\begin{equation}
W=\Theta(|J_{2}|-|J_{1}|),\label{eqn:Wssh}
\end{equation}
where $\Theta$ is the Heaviside function ($\Theta(x)=1$ for $x>0$
and $\Theta(x)=0$ for $x<0$). In order to compute the winding number
in real space for the non-Hermitian SSH model, we follow the steps
described in Sec.~\ref{sec:real_space_method}. First, the Hamiltonian
is written in the order of sublattices and divided into four blocks,
as in Eq.~\eqref{eqn:Hr}. In this case, $\Lambda=H_{1,1}=\epsilon_{A}\leavevmode{\rm 1\ifmmode\mkern-4.8mu \else\kern-0.3em \fi I}$, and $V=H_{1,2}$. From Eq.~\eqref{eqn:Hr}, we get $U\tilde{V}=V$,
where $U$, $V$ and $\tilde{V}$ are all of dimension $M\times M$.
To determine the unitary matrix $U$, we need to fulfill two requirements:
i) the columns of $U$ need to be eigenvectors of $\Lambda$ and
ii) $\tilde{V}$ needs to be positive definite. Since $\Lambda\propto\leavevmode{\rm 1\ifmmode\mkern-4.8mu \else\kern-0.3em \fi I}$,
the first requirement is satisfied for any vector. In order to satisfy
the second requirement, we recall that the polar decomposition of
an invertible square matrix $V$ is a factorization of the form $V=U\tilde{V}$,
where $U$ is a unitary matrix and $\tilde{V}$ is a positive-definite
Hermitian matrix. $\tilde{V}$ is uniquely determined by
$\tilde{V}=(V^{\dagger}V)^{1/2}$. As a result, $U$ can be written
as 
\begin{equation}
U=V(V^{\dagger}V)^{-1/2}.
\end{equation}
%where $V=H_{AB}$. 

Finally, the winding number $W$ can be calculated via Eq.~\eqref{eqn:Wr}. From here on, we set the on-site potentials to be zero, i.e., $\epsilon_{A}=\epsilon_{B}=0$.

\begin{figure}
\begin{centering}
\includegraphics[scale=0.5]{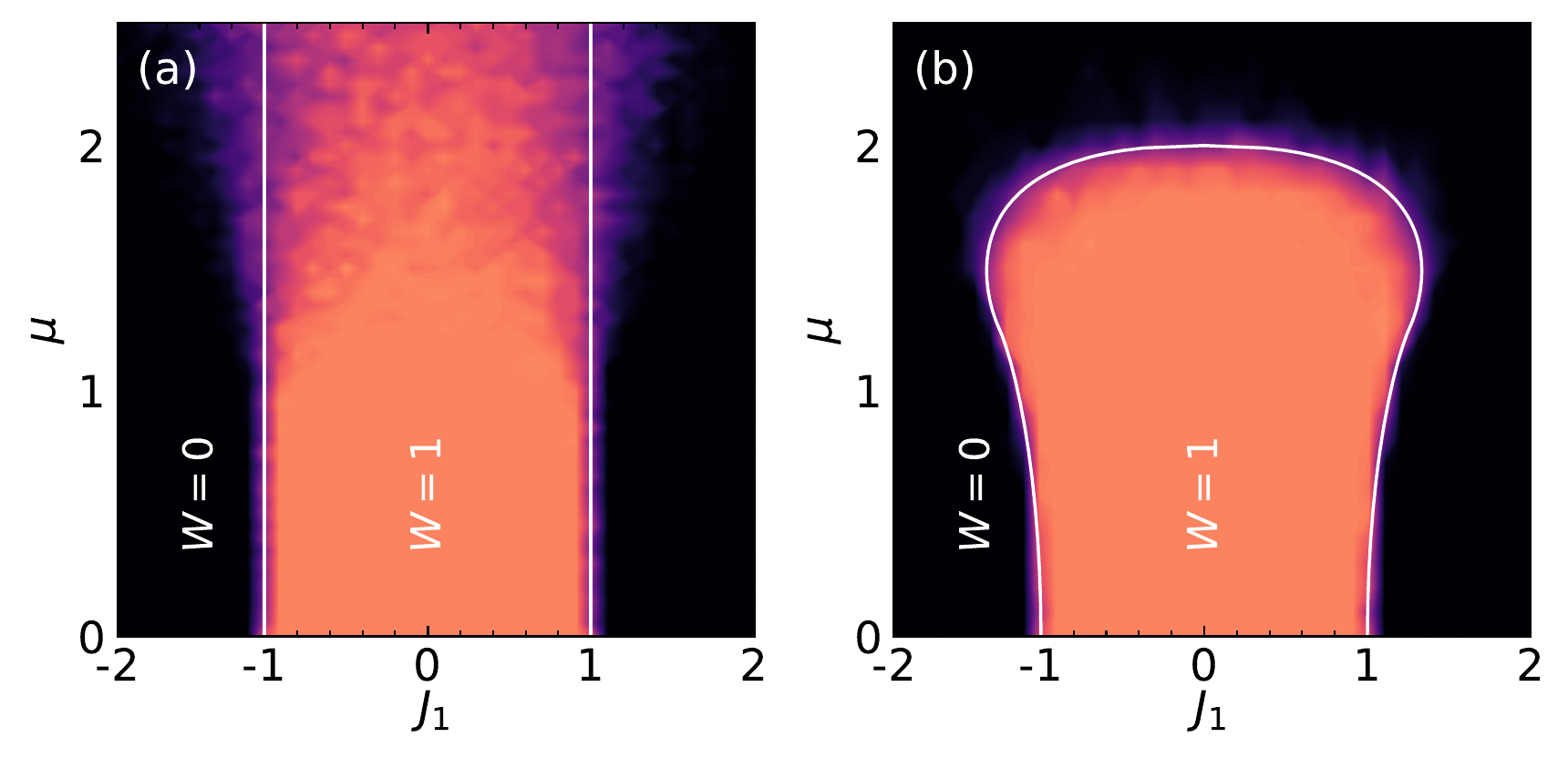} \caption{Phase diagram of the disordered, dissipative SSH model for $N=1000$, $\Gamma=0.5$ and $J_2=1$. Results are averaged over 40 random realizations.
(a) isotropic disorder ($\mu_{1}=\mu_{2}=\mu$), (b) anisotropic disorder
($\mu_{1}=2\mu_{2}=\mu$). White lines indicate points of diverging
localization length in the thermodynamic limit. \label{fig:pd_ssh}}
\par\end{centering}
\end{figure}
Fig.~\ref{fig:pd_ssh} presents the phase diagrams of the disordered
dissipative SSH model as a function of coupling and disorder strength.
In Fig.~\ref{fig:pd_ssh}~(a), the disorder is isotropic, i.e.~$\mu_{1}=\mu_{2}=\mu$
and $J_{2}=1$, while in Fig.~\ref{fig:pd_ssh}~(b),
we consider anisotropic disorder with $\mu_{1}=2\mu_{2}=\mu$ and $J_{2}=1$. The
exact location of the phase transition, illustrated by the white lines
in Fig.~\ref{fig:pd_ssh}, can be obtained analytically by studying
loci of the divergences in the localization length of the edge modes
\citep{Mondragon2014,book:Palyi2016}, as elucidated in more detail in the
Appendix, Sec.~\ref{sec:analytic_contour}. In Fig.~\ref{fig:pd_ssh}~(a),
the phase transition occurs at $|J_{2}/J_{1}|=1$ for all disorder
strengths as for the clean case. %However, one can notice the numerically obtained phase boundary getting fuzzy at large disorder for finite sized chains. In contrast, 
Fig.~\ref{fig:pd_ssh}~(b) shows a non-trivial \emph{topology by disorder} effect. Namely, for
fixed value of $|J_{1}|>1$ close to one, one enters the topologically
non-trivial region by increasing the disorder strength $\mu$, before
transitioning into the topologically trivial regime after further
increasing the noise. This widening of the topological phase boundary
is observed for any kind of anisotropic disorder $\mu_{1}\neq\mu_{2}$.
%, occurring in the weak (strong) disordered regime for $\mu_2<\mu_1$ ($\mu_2>\mu_1$). 

%In Appendix, Section \ref{sec:appendix}, the dependence of the phase boundary on the disorder parameters is derived in the small disorder limit. Here, we  discuss this widening effect on a more intuitive level. 

As already mentioned, the exact phase transition points can be evaluated
from the divergence of the localization length.
In particular, the
phase boundary of the disordered SSH model is given by the equation
$\mathbb{E}(\log|J_{1,j}|)=\mathbb{E}(\log|J_{2,j}|)$, where $\mathbb{E}(\bullet)$
denotes average over disorder (see Eq.~\eqref{eq:loc_ssh}). We first discuss the widening at small
disorder strengths observed in Fig.~\ref{fig:pd_ssh}~(b). The second order
Taylor expansion of $\mathbb{E}(\log|X|)$ in $\mu_{i}/J_{i}$ reads $\mathbb{E}[\log|X|]\simeq\log(\mathbb{E}[X])-\frac{\mathbb{V}[X]}{2\mathbb{E}[X]^{2}}$ \citep{NIPS2006_3113},
resulting in the following approximation of the phase boundary equation,

\begin{equation}
\log|J_{1}|-\frac{\mu_{1}^{2}}{6J_{1}^{2}}\simeq\log|J_{2}|-\frac{\mu_{2}^{2}}{6J_{2}^{2}}.\label{eq:phaseb_approx}
\end{equation}

Fixing $J_{2}$ and $\mu_{2}$ such that the right hand side of
Eq.~(\ref{eq:phaseb_approx}) is constant, we see that the function $\log|J_{1}|-\frac{\mu_{1}^{2}}{6J_{1}^{2}}$
is monotonically increasing in $J_{1}$ and decreasing in $\mu_{1}$.
Hence, if $\mu_{1}$ increases, $J_{1}$
needs to grow as well in order to compensate. This corresponds to a widening
of the topologically non-trivial region for small increasing noise.
In the opposite, strong disorder limit, we can expand $\mathbb{E}(\log|J_{\alpha,i}|)$
in $J_{\alpha}/\mu_{\alpha}$, obtaining $\mathbb{E}(\log|J_{\alpha,i}|)=\log|\mu_{\alpha}|-1+O(J_{\alpha}/\mu_{\alpha})$.
The phase boundary equation in this regime becomes

\begin{equation}
\log|\mu_{1}|\simeq\log|\mu_{2}|.
\end{equation}
Hence, for strong disorder, the phase boundary is roughly independent
of $J_{1}$ accounting for the horizontal boundary in
Fig.~\ref{fig:pd_ssh}~(b). %the disorder in hopping dominates over the original finite hopping, leading to the fact that the expected value is merely determined by the strength of disorder. As a result, the system 
%enters the non-trivial phase whenever the disorder in $J_{2j}$ is larger than the disorder in $J_{1j}$, regardless of the initial configurations. 
Similar disorder-induced topological characteristics were also recently
discussed in the context of other non-Hermitian models \citep{Luo2019,Zhang2020}. \\ 

\begin{figure}
\begin{centering}
\includegraphics[scale=0.53]{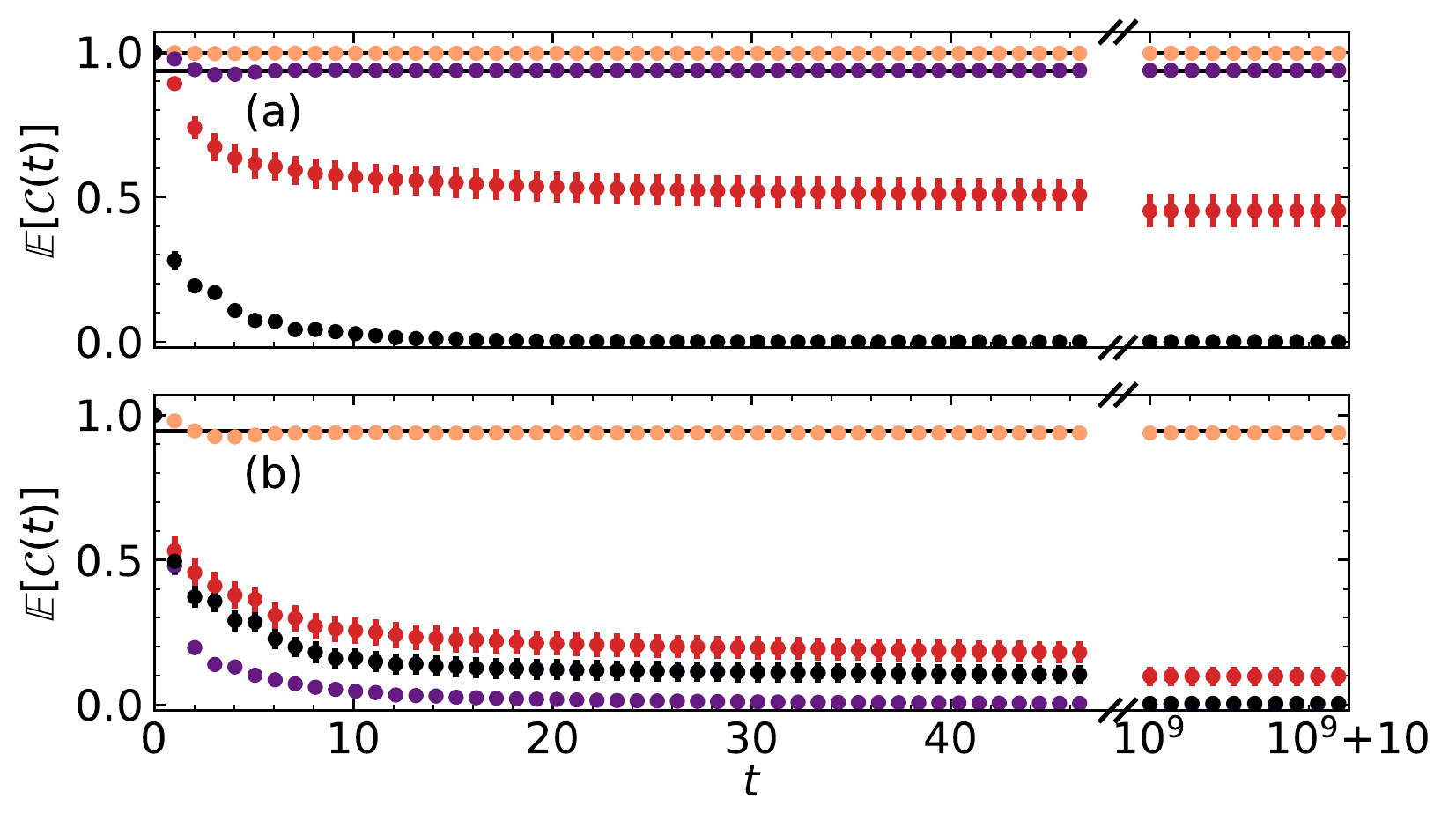} \caption{Coherence of the first qubit in the disordered, dissipative SSH model for $N=100$, $\Gamma=0.5$ and $J_2=1$. Results are averaged over 40 random realizations. 
(a) isotropic disorder ($\mu_{1}=\mu_{2}=\mu$) along the vertical line where $J_1=0$ with $\mu=0.1$ (orange), $\mu=0.5$ (purple), $\mu=1.0$ (red), and in the topologically trivial regime $J_1=1.5$ with $\mu=0.5$ (black). (b) anisotropic disorder
($\mu_{1}=2\mu_{2}=\mu$) along the vertical line where $J_1=1.2$ with $\mu=0.5$ (purple), $\mu=1.5$ (red), $\mu=2.5$ (black), and for $J_1=1.5$ with $\mu=0.5$ (orange). Solid black lines indicate the asymptotic prediction $\mathbb{E}(1-x^2)$ Eq.~(\ref{eq:cinf_approx}) valid for small disorders. \label{fig:coh_ssh} }
\par\end{centering}
\end{figure}

For each phase diagram, we now fix $J_2=1$ and choose four characteristic parameter configurations
in order to get representative coherence time evolutions for the different
topological sectors, depicted Fig.~\ref{fig:coh_ssh}.
For isotropic disorder, Fig.~\ref{fig:coh_ssh}~(a), we choose three points along the vertical $J_1=0$ with $\mu=0.1,0.5,1.0$ as well as the configuration $J_1=1.5, \mu=0.5$, representing the disordered topologically non-trivial and trivial regime, respectively.
The coherence decays to a non-zero (respectively zero) value at large
times in the topologically non-trivial (respectively
trivial) sector, thus matching the phase diagram Fig.~\ref{fig:pd_ssh}~(a). In the topologically non-trivial regime, increasing disorder leads to a smaller asymptotic value of the coherence. 
Similarly, for anisotropic disorder, Fig.~\ref{fig:coh_ssh}~(b), we choose three points along the vertical $J_1=1.2$ with $\mu=0.5,1.5,2.5$ as well as $J_1=0,\mu=0.5$. The former three parameter pairs lie on a vertical line cutting through the broadening of the topologically non-trivial regime, thus representing the reentrance phenomenon into a higher topological phase. It can be seen that a finite coherence of the first qubit is present at large times only for $\mu=1.5$, being in consent with the corresponding phase diagram Fig.~\ref{fig:pd_ssh}~(b). For $J_1=0$ and $\mu=0.5$, a similar behavior as for the isotropic disordered chain can be observed, with a large asymptotic coherence value. 
In previous work \citep{Venuti2017}, it was shown that for large chains in the topologically non-trivial regime, the coherence saturates to approximately
\[
\mathcal{C}(t\rightarrow\infty)\approx 1-x^2,
\]
where $x=J_1/J_2$, with $|x|<1$. It is thus natural to assume that the expectation value of the asymptotic coherence including disorder is given by  
\begin{gather}
\mathbb{E}[\mathcal{C}(t\rightarrow\infty)]\approx \mathbb{E}(1-x^2) = \nonumber \\ \frac{1}{4\mu_1 \mu_2} \int_{-\mu_2}^{\mu_2}\int_{-\mu_1}^{\mu_1} 1- \Big( \frac{J_1+\mu_1}{J_2+\mu_2} \Big)^2 d\mu_1 d\mu_2 = \label{eq:cinf_approx}  \\ 1-\frac{3J_1^2 + \mu_1^2}{3J_2^2-3\mu_2^2} \, . \nonumber
\end{gather}
Note that this only holds for weak to moderate disorder such that no change of topological phase can be generated randomly, i.e., $\mu_1 + \mu_2 < |J_2| - |J_1|$.
%\begin{equation}
%\mu_1 + \mu_2 < 
%\begin{cases}
%\, \, \, \, \, J_2 - J_1 \qquad J_1>0, J_2>0 \\
%-J_2 - J_1 \qquad J_1>0, J_2<0 \\
%- J_2+ J_1 \qquad J_1<0, J_2<0 \\
%\, \, \, \, \, J_2 + J_1 \qquad J_1<0, J_2>0.
%\end{cases}
%\end{equation}
In Fig.~\ref{fig:coh_ssh}, the prediction Eq.~(\ref{eq:cinf_approx}) is illustrated by black solid lines for disorder strengths falling into the discussed regime. For large disorder, random phase changes result in a decrease of the mean coherence in the simulation, and Eq.~(\ref{eq:cinf_approx}) breaks down.
\\ \\
It is important to note that, even though the phase diagram is the
same as those found in previous works \citep{Mondragon2014,Luo2019},
the physical interpretation is different, as our models include dissipation.
Edge states do not correspond to actual electronic states located
at one of the boundaries of the chain, but rather describe the physics
of the projected density matrix introduced in Sec.~\ref{sec:setup}.
A non-trivial topological phase, resulting in quasi-dark states of
the restricted Hamiltonian, leads to having an exponentially long
(in system size) coherence time of the edge qubit. In the topologically
trivial regime, the decoherence of the edge qubit is governed by dissipation,
leading to a finite coherence time.

\subsection{Disordered non-Hermitian Trimer Model}

\label{sec:3site}

\begin{figure}
\centering{}\includegraphics[width=0.48\textwidth]{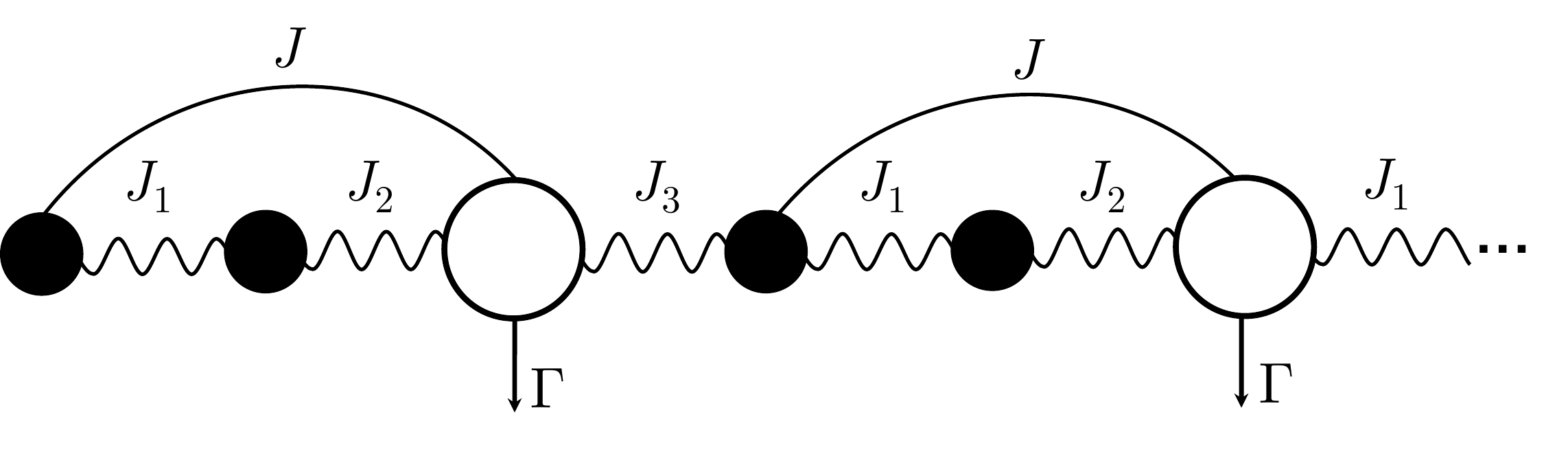} \caption{Non-Hermitian trimer model. Here, the nearest-neighbor couplings $J_{1},J_{2},J_{3}$
alternate cyclically, building a unit cell with three sites. Next-nearest-neighbor
couplings $J$ link the first and third site in each unit cell, thus
enabling three distinct winding numbers $W=0,1,2$.}

\label{fig:3site} 
\end{figure}

Next, we consider a trimer chain with nearest-neighbor as well as
next-nearest-neighbor couplings, as depicted in Fig.~\ref{fig:3site}.
The corresponding non-Hermitian Hamiltonian, derived from the restricted
Lindbladian, is given by 
\begin{eqnarray}
\mathsf{H} & = & \sum_{j=1}^{M}(J_{1,j}|j,B\rangle\langle j,A|+\text{h.c.})\nonumber \\
 & + & \sum_{j=1}^{M-1}(J_{2,j}|j,C\rangle\langle j,B|+\text{h.c.})\nonumber \\
 & + & \sum_{j=1}^{M-1}(J_{j}|j,C\rangle\langle j,A|+\text{h.c.}) \\
 & + & \sum_{j=1}^{M-1}(J_{3,j}|j+1,A\rangle\langle j,C|+\text{h.c.})\nonumber \\
 & + & \sum_{j=1}^{M}\epsilon_{A}|j,A\rangle\langle j,A|+\sum_{j=1}^{N}\epsilon_{B}|j,B\rangle\langle j,B|\nonumber \\
 & + & \sum_{j=1}^{M}(\epsilon_{C,j}-i\Gamma)|j,C\rangle\langle j,C|.\label{3site_H}\nonumber
\end{eqnarray}
%\begin{equation*}
%\mathsf{H}=\left(\begin{array}{ccccccc}%{*7{C{3.5em}}} %put this instead of {cccccc} for equal spacing of columns
%\epsilon_{A,1} & J_{1,1} & J_1 & 0 & 0 & 0  & 0 \\
%J_{1,1} & \epsilon_{B,1} & J_{2,1} & 0 & 0 & 0& 0 \\
%J_1 & J_{2,1} & \epsilon_{C,1} - i\Gamma & J_{3,1} & 0 &0& 0 \\
%0 & 0 & J_{3,1} & \epsilon_{A,2} & J_{1,2} & J_2 & 0 \\
%0 & 0 & 0 & J_{1,2} & \epsilon_{B,2} & J_{2,2} & \ddots \\
%0 & 0 & 0& J_2 & J_{2,2} & \epsilon_{C,2}-i\Gamma & \ddots \\
%0 & 0 & 0 & 0 & \ddots & \ddots & \ddots
%\end{array}\right). \label{eq:H_SSH}
%\end{equation*}
It has been demonstrated that robust chiral edge modes exist in non-dissipative trimer chains, 
even in the absence of inversion symmetry 
\citep{Alvarez2019}. It has been argued that their topological character is inherited through a mapping of a higher-dimensional model, namely the commensurate off-diagonal Aubry-André-Harper  model, which is topologically equivalent to a two dimensional tight-binding lattice pierced by a magnetic flux \citep{Kraus2012}.
%Note that even in the non-dissipating, clean trimer chain ($J_{1,j}=J_{1},J_{2,j}=J_{2},J_{3,j}=J_{3},J=0,\Gamma=0$),
%no symmetries are present which would \textit{a priori} predict non-trivial
%topology and robust edge states in the system. Therefore, according
%to the standard classification of topological insulators no winding
%number can be defined \citep{Schnyder2008}. Nonetheless, it has been
%argued that the existence of topological edge states in trimer chains
%is inherited through a mapping of a higher-dimensional model, namely
%the commensurate off-diagonal Aubry-André-Harper model \citep{Alvarez2019},
%which is topologically equivalent to a two dimensional tight-binding
%lattice pierced by a magnetic flux \citep{Kraus2012}, hence being
%of topological nature and hosting the quantum Hall effect. \textbf{I'm
%not sure what you're saying here. Ref~}\citep{Schnyder2008} \textbf{concerns
%only 3D. In 1D you }\textbf{\emph{need}}\textbf{ a symmetry in order
%to have non-trivial topology or else every model is topologically
%trivial. }
The topological classification by Rudner \textit{et al.}~including
dissipation, however, imposes only translational symmetry. In fact,
it turns out that the winding number in Eq.~\eqref{eqn:Wr} can be
used as a reliable predictor for the number of (quasi)-dark states
located on the edge of the trimer chain with open boundary conditions.
In previous work \citep{Venuti2017}, it was found that in the clean
case, the presence of next-nearest-neighbor couplings enable winding
numbers $W=0,1,2$. Concretely, $W$ is given by %For the model under current consideration, it is easy to see that $v_k = (J+J_3 e^{ik}, J_2)^T$. Choosing the phases $z$ in order to fulfill the two criteria for $U$, the winding number breaks down to the winding number of $J+J_2+ J_3 e^{ik}$ plus the one of $J - J_2 + J_3 e^{ik}$. Therefore, the next nearest neighbor hopping enables winding numbers $W=0,1,2$. Concretely, $W$ is given by 
\begin{align}
W & =\Theta\left(\left|J_{3}\right|-\left|J+J_{2}\tan(\vartheta/2)\right|\right)\nonumber \\
 & +\Theta\left(\left|J_{3}\right|-\left|J-J_{2}\cot(\vartheta/2)\right|\right),\label{windingeq_3site}
\end{align}
where $\vartheta=\arccos\left[\left(\epsilon_{A}-\epsilon_{B}\right)/\sqrt{4J_{1}^{2}+\left(\epsilon_{A}-\epsilon_{B}\right)^{2}}\right].$
We further verify the above equation in the Appendix, Sec.~\ref{sec:appendix}, by solving the
system analytically for a convenient system size and counting the
number of dark states localized on one edge of the chain.
\begin{figure}
\begin{centering}
 \includegraphics[scale=0.48]{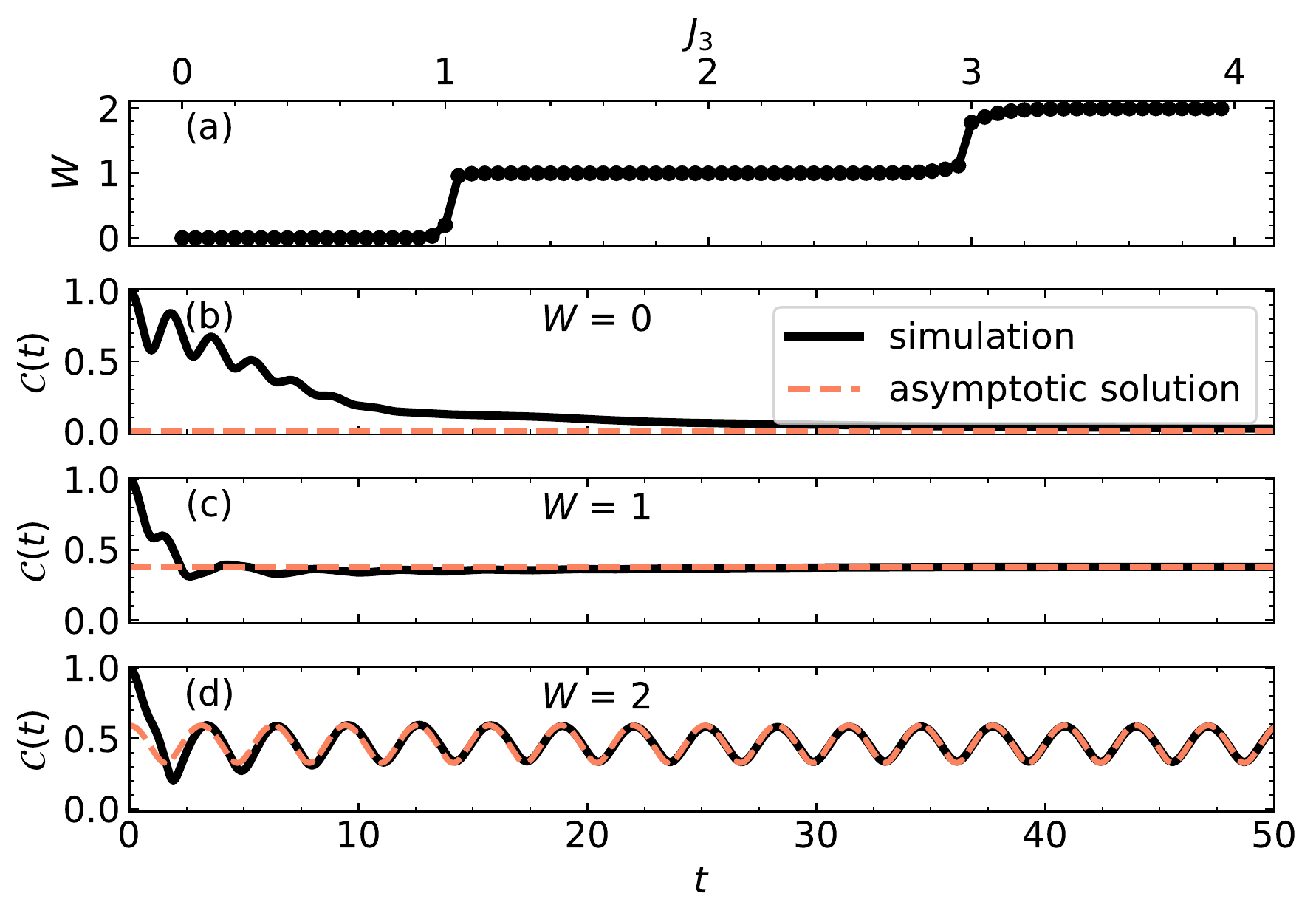} \caption{Topology and coherence for the clean, non-disordered trimer model. (a) Winding number and (b)-(d) time dependent coherence for three
parameter configurations corresponding to the three topological sectors.
The dotted lines indicate the theoretically predicted asymptotic coherence
as $t\rightarrow\infty$. Dissipation is set to $\Gamma=0.5$ and
a chain with $N=300$, $J_{1}=1,J_{2}=2$ and $J=1$ is considered.
The three time evolutions of the coherence in the topological sectors
$W=0,1,2$ correspond to parameter choices $J_{3}=0.5,2.0,3.5$, respectively. \label{fig:clean_3site}}
\par\end{centering} 
\end{figure}
In order to calculate the winding number using the the real-space approach, 
we first rewrite the Hamiltonian with respect to its sublattices
and decompose it as in Eq.~\eqref{eqn:Wr}. In this case, the matrices
$\Lambda$ and (respectively $V$) with dimensions $2M\times2M$ 
(respectively $2M\times M$) are given by 
\begin{equation}
\Lambda=\begin{pmatrix}\epsilon_{A}\mathbb{1} & H_{AB}\\
H_{BA} & \epsilon_{B}\mathbb{1}
\end{pmatrix} ;\qquad V=\begin{pmatrix}H_{AC}\\
H_{BC}
\end{pmatrix}.
\end{equation}
Here, $H_{AB}=J_{1}\mathbb{1}$. Due to the symmetry of $\Lambda$,
$U$ from Eq.~\eqref{eqn:Hr} can be written as 
\begin{equation}
U=\begin{pmatrix}-\cos(\vartheta/2)U_{-} & \sin(\vartheta/2)U_{+}\\
\sin(\vartheta/2)U_{-} & \cos(\vartheta/2)U_{+}
\end{pmatrix},\label{eq:U_vsUpm}
\end{equation}
where $U_{\pm}$ are two $M\times M$ so far unspecified unitaries
and $\vartheta$ has been given above. From Eq.~\eqref{eqn:Hr}, we
further get $U\tilde{V}=V$, which gives 
\begin{equation}
\begin{pmatrix}-\cos(\vartheta/2)U_{-} & \sin(\vartheta/2)U_{+}\\
\sin(\vartheta/2)U_{-} & \cos(\vartheta/2)U_{+}
\end{pmatrix}\begin{pmatrix}\tilde{V}_{-}\\
\tilde{V}_{+}
\end{pmatrix}=\begin{pmatrix}H_{AC}\\
H_{BC}
\end{pmatrix},
\end{equation}
where $\tilde{V}:=(\tilde{V}_{-},\tilde{V}_{+})^{T}$. From the above
equation we find 
\begin{align}
U_{+}\tilde{V}_{+} & =\frac{1}{2}(\cos(\vartheta/2)H_{AC}+\sin(\vartheta/2)H_{BC}),\nonumber \\
U_{-}\tilde{V}_{-} & =\frac{1}{2}(-\sin(\vartheta/2)H_{AC}+\cos(\vartheta/2)H_{BC}).\label{eq:3site_decomp}
\end{align}
Recall that we must fix the gauge freedom in $U$ by requiring the
submatrices $\tilde{V}_{\pm}$ to be positive definite. Consequently,
$U_{\pm}$ can be determined by polar decomposition of the right hand
side of Eq.~\eqref{eq:3site_decomp}, after which the unitary matrix
$U$ is obtained using Eq.~\eqref{eq:U_vsUpm}. Finally, the winding
number is computed via Eq.~\eqref{eqn:Wr}. Interestingly, it can
be shown that the winding number of $U$ is nothing more than the
sum of the winding numbers of $U_{+}$ and $U_{-}$.

\begin{figure}[t]
\begin{centering}
\includegraphics[scale=0.5]{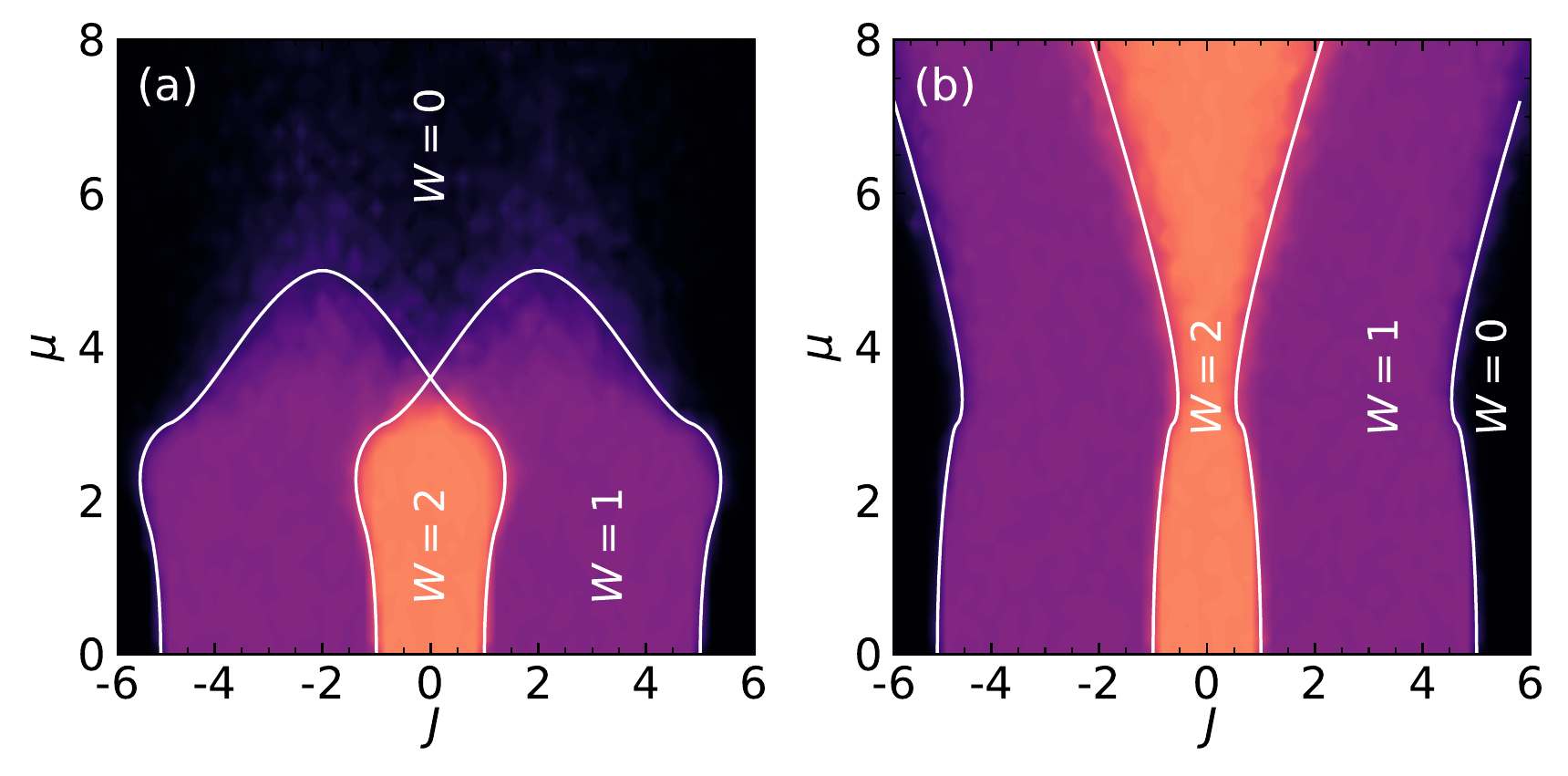} \caption{Phase diagram of the disordered, dissipative trimer model for $N=1500$, $\Gamma=0.5$
, $J_1=2$, $J_2=2$, $J_3=3$. Results are averaged over 40 random realizations.
In (a), $\mu_{2}=\mu_{J}=\mu_{3}=\mu$, whereas (b) describes disorder
with $2\mu_{J}=2\mu_{2}=\mu_{3}=\mu$. White lines indicate
the loci of diverging localization lengths in the thermodynamic limit. \label{fig:pd_3site} }
\par\end{centering}
\end{figure}

For simplicity, we again limit our considerations to the case of
vanishing the on-site chemical potentials, i.e., $\epsilon_{A}=\epsilon_{B}=\epsilon_{C}=0$. 
Using the real-space winding number approach for the clean trimer model results in Fig.~\ref{fig:clean_3site}~(a),
matching Eq.~\eqref{windingeq_3site}. Figs~\ref{fig:clean_3site}~(b)-(d) show the typical behavior of the coherence in the three distinct topological sectors $W=0,1,2$ in the clean trimer model, respectively. In the topologically trivial regime, no dark states are present, driving decoherence of the first qubit. For $W=1$, the dark state manifold is one-dimensional, leading to a saturation of the coherence at infinite times. For $W=2$, the existence of two dark states result in Rabi like oscillations of the first qubit's coherence. The asymptotic solution, Eq.~\eqref{theo_coh}, is also featured in Figs.~\ref{fig:clean_3site}~(b)-(d). 
%Fig.~\ref{fig:clean_3site}(b)
%illustrates how in the topologically trivial regime, dissipation again
%leads to decoherence of the edge qubit, whereas non-trivial topology
%results in an exponentially large coherence time \citep{Venuti2017}. For $W=1$,
%only one dark state contributes to the evolution of the coherence,
%as the overlap of the left edge site with the dark state located on
%the right edge of the chain is exponentially small. For $W=2$, on
%the other hand, both dark states are located on the side of the considered
%qubit, resulting in Rabi-like oscillations of the coherence at infinite
%times, cf. Eq.~\eqref{theo_coh}. 
Because of the $J_{1}$ dependence
of the dark states, disorder in $J_{1}$ is expected to quickly destroy the
topological features of the system. This is further suggested
by the degree of freedom of the matrix $U$, Eq.~\eqref{eqn:Hr}, which
collapses as soon as $J_{1}$ becomes disordered, leading to an immediate
collapse of a well-defined winding number. Therefore, we shall from now on focus
on the analysis of the disordered regime where only $J_{2},J_{3},J$ are exposed to noise, 
which we control via additive random noise drawn from a uniform distribution. Concretely,
if $j$ labels the unit cell and $\{\omega_{1}\}$, $\{\omega_{2}\}$,
$\{\omega\}$ are sets of independent, uniformly distributed random
variables $\in[-1,1]$, $J_{i,j}=J_{i}+\mu_{i}\omega_{i,j}$ for $i=2,3$,
$J_{j}=J+\mu_{J}\omega_{j}$, and $J_{1,j}=J_{1}$ for all $j$. Looking
at the density of states for the different disorder types, depicted
in Figure \ref{fig:trimer_dos}, the selection rules
for the type of disorder under which topological dark states are stable
is further underlined.

As for the disordered non-Hermitian SSH model, the full phase diagram
for different disorder strengths can be constructed, shown in Fig.~\ref{fig:pd_3site}.
Again, the exact phase transition
points in the thermodynamic limit are depicted by white lines, which are derived 
via the dark state localization length considering disorder in 
the Appendix, Sec.~\ref{sec:appendix}. The phase
diagram features rich structures, presenting widenings of topologically
non-trivial phases for moderate (high) disorder strengths in the chain
with equal (different) disorder amplitudes. Note that the system with
different distributions on the disordered parameters, $2\mu_{2}=2\mu_{J}=\mu_{3}=\mu$,
is more similar to what we called anisotropic disorder in the SSH model, being due 
to the competition between $|J_{2}\pm J|$ and $J_{3}$
deciding the topological phase for the trimer model Eq.~\eqref{windingeq_3site}.
When computing the localization length, the disorder amplitudes of
$J_{2}$ and $J$ hence add up, as is explicitly seen in Eq.~\eqref{eq:3siteLL}.
Note, however, that the effective disorder on $|J_{2}\pm J|$ is $\mu/\sqrt{2}<\mu$,
which results in having a widening of the non-topological phases in
the large disorder regime. Analogously, the trimer system having equal
disorder on all hopping parameters resembles the case $\mu_{1}>\mu_{2}$
of the SSH-model, featuring a widening of the topologically non-trivial
regimes for small disorders. \\ \\ 
\begin{figure}[t]
\begin{centering}
\includegraphics[scale=0.53]{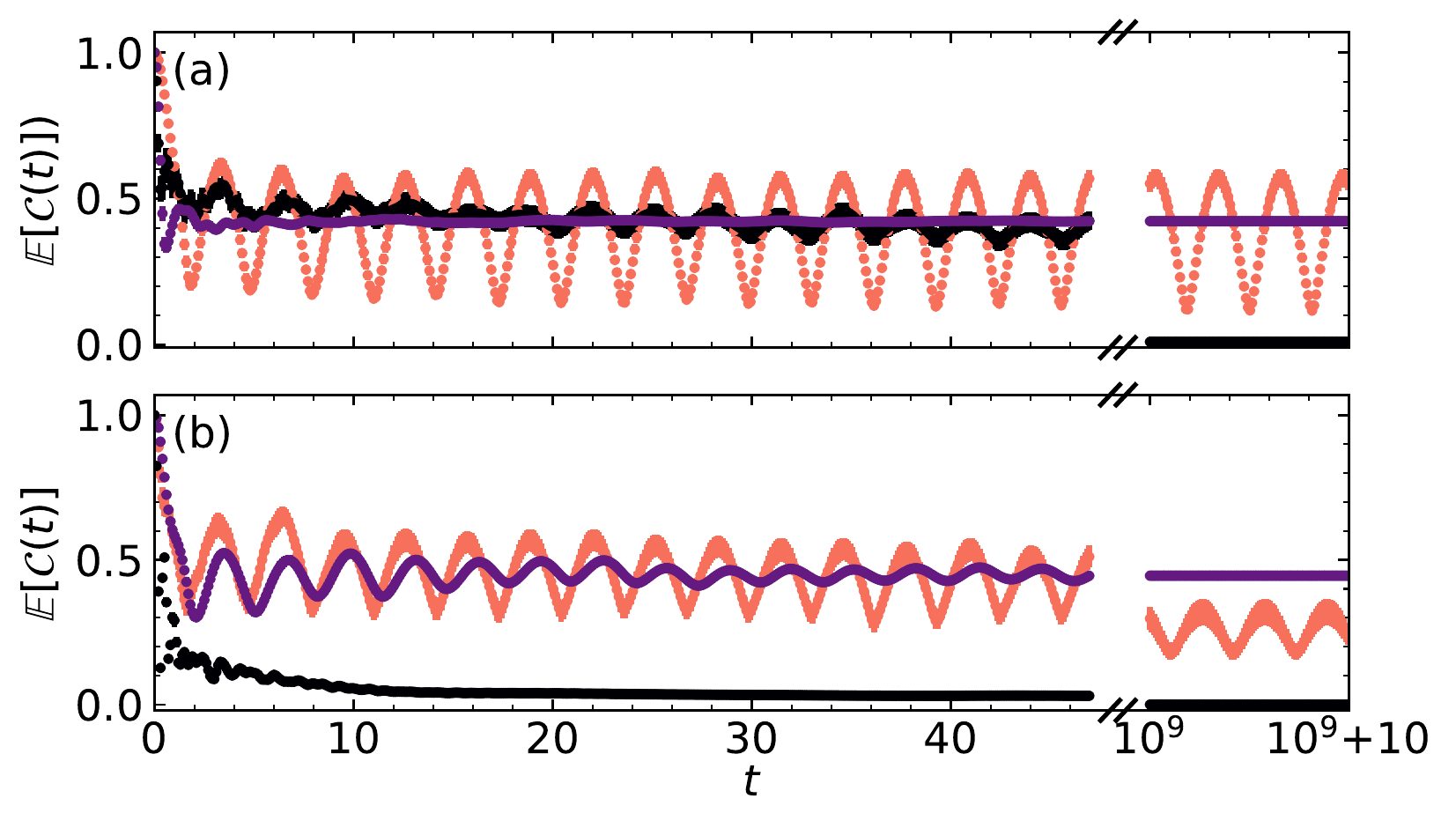} \caption{Coherence of the first qubit in the disordered, dissipative trimer model for $N=300$, $\Gamma=0.5$, $J_1=1$, $J_2=2$ and $J_3=3$. Results are averaged over 40 random realizations.
(a) For $\mu_{J}=\mu_{2}=\mu_{3}=\mu$, we show the coherence for $(J,\mu)=(0,1)$ (orange), $(J,\mu)=(3,1)$ (purple), and $(J,\mu)=(0,7)$ (black), corresponding to $W=2,1,0$, respectively. (b) For $2\mu_{J}=2\mu_{2}=\mu_{3}=\mu$, we highlight the reentrance into a higher topological phase along the vertical line $J=1.2$, with $\mu=1$ (purple) and $\mu=7$ (orange), corresponding to $W=1,2$, respectively. We further show the trivial regime by evaluating the coherence for $(J,\mu)=(6,1)$ (black). The observable broadening of the curves is due to the error of the mean, pictured by error bars for every data point. \label{fig:coh_3site} }
\par\end{centering}
\end{figure}
We shall again pick three
points in each phase diagram and illustrate the corresponding time evolution of the first 
qubits coherence, seen in Fig.~\ref{fig:coh_3site}. For $\mu_{J}=\mu_{2}=\mu_{3}=\mu$, Fig.~\ref{fig:coh_3site}~(a), 
we choose the parameter pairs $(J,\mu) = (0,1), (3,1), (0,7)$, belonging to winding numbers 
$W=2,1,0$, respectively (cf. Fig.~\ref{fig:pd_3site}). For all configurations, we find that the 
asymptotic behavior of the coherence the one of the clean case, namely a decrease 
to zero for $W=0$, a convergence to a constant larger than zero for $W=1$, and an oscillation 
for $W=2$. For different disorder strengths $2\mu_{J}=2\mu_{2}=\mu_{3}=\mu$, we focus on 
the reentrance phenomenon $W=1\rightarrow 2$ by computing the coherence for 
$(J,\mu)=(1.2,1), (1.2,7)$. Indeed, we find that for large enough disorder, an oscillating behavior emerges, signaling the change of topological phase. For completeness, we also include $(J,\mu)=(6,1)$ representing the trivial sector, where a vanishing coherence can be observed at large times.

\section{Application to Quantum Computation}
\label{sec:quantum_computation}
Ever since Kitaev's proposal \citep{Kitaev2006} to braid anyons in
order to realize non-trivial quantum gates, the field of topological
quantum computation has been an exceptionally active field of research
\citep{Sankar2008, Sau2010, Sarma2015, Stern2013, Alicea2011, Freedman2006, Freedman2003, Akhmerov2010}. This is mainly due to the
promising protection against environmental noise governed by the non-locality
of the state manifold used for braiding \citep{Lahtinen2017}. Spinless
p-wave superconductor wires hosting non-Abelian Majorana fermions
bound to topological defects have been of particular interest \citep{Fisher2010},
as the intrinsic particle-hole symmetry of the BdG-Hamiltonian promises
a realizable topological protection. Recently, the SSH model has been
analyzed in terms of its applicability to quantum computation \citep{Andras2019},
where it was found that the non-trivial braiding statistics of the
topological edge modes can be used to build quantum gates via Y-junctions.
However, as for all quantum gates based on symmetry protected topological
states, the set of quantum gates is not universal \citep{Lahtinen2017}.
Nevertheless, studying the braiding statistics for our concrete open
disordered models seems like an exciting and promising work for future
projects.

\section{Conclusions}
\label{sec:conclusion}
We have analyzed and topologically classified disordered dissipative qubit-cavity dimer and trimer architectures, with special focus on topological protection mechanisms of the coherence measure in a fiducial qubit. The evolution of the coherence's qubit is exactly given by a non-Hermitian Hamiltonian which thus emerges from a bona-fide physical system. 
We demonstrated the use of a real-space topological invariant $W$, which accurately predicts the number of non-trivial (quasi-)dark modes in disordered, non-Hermitian models, as long as certain symmetries are preserved by the disorder operators. We then computed the phase diagrams of dimer and trimer chains in the parameter space spanned by the tunneling amplitude and the disorder strength,  predicting the faith of the fiducial qubit's coherence at long times, i.e., decay to zero, a constant value or oscillatory behavior for winding numbers $W=0,1,2$, respectively. For certain choices of disorder strengths or the hopping parameters, reentrance phenomena into topological phases with higher winding numbers were observed, leading to an increase of coherence times (exponentially large in system size) when introducing higher noise levels. Possible applications in topological quantum computing via braiding of dark modes were briefly discussed, opening up interesting questions for future research. Furthermore, generalizations of the classification to larger numbers of sites per unit cell and systems of higher dimension would be of great interest. 

\textbf{Acknowledgements:} We would like to thank Hubert Saleur for useful discussions.
This work was supported by the US Department of Energy under grant
number DE-FG03-01ER45908. L.C.V. acknowledges partial support from the
Air Force Research Laboratory award no. FA8750-18-1- 0041. The research is based upon work (partially) supported by the Office of the Director of National Intelligence (ODNI), Intelligence Advanced Research Projects Activity (IARPA), via the U.S. Army Research Office contract W911NF-17-C-0050. The views and conclusions contained herein are those of the authors and should not be interpreted as necessarily representing the official policies or endorsements, either expressed or implied, of the ODNI, IARPA, or the U.S. Government. The U.S. Government is authorized to reproduce and distribute reprints for Governmental purposes notwithstanding any copyright annotation thereon. 

\bibliographystyle{apsrev4-1}
\bibliography{open_disordered_topology}
\onecolumngrid
\clearpage 

\appendix

\section{Dark states in the dissipative trimer model}

\label{sec:appendix} 

\label{ds3site} 
We here derive an exact form of the asymptotic coherence dynamics and
the topological phase transition in the trimer model by studying the
dark states, i.e., by finding all states that obey $\mathsf{H}|\psi\rangle\!\rangle=E|\psi\rangle\!\rangle$
with $E\in\mathbb{R}$. For the sake of convenience, the following
considerations assume chain lengths $N\mod3=2$, as the system
then hosts exact dark states with vanishing imaginary part. For all
other system sizes the states
are quasi-dark, as they have an imaginary part exponentially small
in the system size. Of course, in the thermodynamic limit, these differences
vanish, and the dynamics is exactly described by the result below.
The ansatz is to look for possible dark states with energies $E=\pm J_{1}$, i.e., to find the kernel of the matrix 
\begin{equation}
\mathsf{H}\mp\mathbb{1}J_{1}=\begin{pmatrix}\mp J_{1} & J_{1} & J & 0 & 0 & 0\\
J_{1} & \mp J_{1} & J_{2} & 0 & 0 & 0\\
J & J_{2} & \mp J_{1}-i\Gamma & J_{3} & 0 & 0\\
0 & 0 & J_{3} & \mp J_{1} & J_{1} & J\\
0 & 0 & 0 & J_{1} & \mp J_{1} & \ddots\\
0 & 0 & 0 & J & \ddots & \ddots
\end{pmatrix}.\label{dseq}
\end{equation}
For $N\mod3=2$, solutions of \eqref{dseq} are of the form 
\begin{eqnarray}
v_{+} & = & \big(1,1,0,-\delta_{+},-\delta_{+},0,(-\delta_{+})^{2},(-\delta_{+})^{2},0,...,(-\delta_{+})^{\frac{N-2}{3}},(-\delta_{+})^{\frac{N-2}{3}}\big)^{T},\nonumber \\
v_{-} & = & \big(1,-1,0,\delta_{-},-\delta_{-},0,\delta_{-}^{2},-\delta_{-}^{2},0,...,\delta_{-}^{\frac{N-2}{3}},-\delta_{-}^{\frac{N-2}{3}}\big)^{T}.\label{ds}
\end{eqnarray}

These solutions are intuitive and analogous to the open SSH model
\citep{Venuti2017}, in the sense that they disappear on all dissipative
sites. The condition $E=\pm J_{1}$ signals the equivalence of the
first two sites of each unit cell up to a sign factor. Eq.~\eqref{ds}
leads to 
\begin{equation}
\delta_{\pm}=\frac{|J\pm J_{2}|}{J_{3}},
\end{equation}
where the sign of the solution is fixed without loss of generality
by assuming $\delta_{\pm}$ to be positive. The winding number classification
is illustrated in the corresponding vectors, as we find zero, one,
or two dark states localized at the outer left qubit for different
topological sectors, i.e., $W=\Theta(J_{3}>|J-J_{2}|)+\Theta(J_{3}>J+J_{2})$.
Taking into account the normalization factor of the solutions, 
\begin{equation}
A_{\pm}^{-2}=2\sum_{k=0}^{\frac{N-2}{3}}\delta_{\pm}^{2k}=2\frac{1-\delta_{\pm}^{\frac{2N-4}{3}}}{1-\delta_{\pm}^{2}},
\end{equation}
the time dependent coherence can be approximated for large times $t\gg1/\Gamma$,
\begin{align}
\mathcal{C}(t) & =|\langle\langle1|e^{-i\mathsf{H}t}|1\rangle\rangle|\approx|e^{-iJ_{1}t}A_{+}^{2}+e^{iJ_{1}t}A_{-}^{2}|\nonumber \\
 & =|A_{+}^{4}+A_{-}^{4}+2A_{+}^{2}A_{-}^{2}\cos2J_{1}t|.\label{theo_coh}
\end{align}

\section{Analytical Determination of Critical Phase Transition Contours}
\label{sec:analytic_contour}

In \citep{Mondragon2014}, the critical
phase transition surface was derived for the Hermitian SSH model,
using the numerical transfer matrix method and level-spacing statistics
analysis. The analytical critical phase transition contour for non-Hermitian
models can be calculated in a similar manner. To see this, consider
the non-Hermitian SSH model. Here, the dark edge state is exactly
at zero energy and only lives on the non-decaying sublattice.
We consider the critical phase transition
in the thermodynamic limit, such that the results for the linear chain
of odd length coincide with the results of even length. Now recall
that the edge state of the disordered Hermitian SSH model is also
supported entirely by one sublattice or the other. Its zero energy
edge state on sublattice A ($\psi_{n,B}=0$) can be written as 
\begin{equation}
\psi_{n,A}=i^{n-1}\prod_{j=1}^{n}\left|\frac{J_{1j}}{J_{2j}}\right|\psi_{1,A},\label{eq:wf_phasetrans}
\end{equation}
where $J_{1}j$ and $J_{2}j$ are the two perturbed hopping parameters
in the $j$th unit cell. The edge states in the two systems share
an identical distribution in the clean limit. Consequently, in this
case, the non-Hermitian problem follows the same localization length
and phase transition as a one-dimensional Hermitian SSH model.

With the exact wave function distribution as in Eq.~\eqref{eq:wf_phasetrans},
the inverse localization length of a edge mode can be obtained by
\begin{eqnarray}
\Lambda^{-1} & = & -\lim_{n\rightarrow\infty}\frac{1}{n}\log\left|\psi_{n,A}\right|\nonumber \\
 & = & \left|\lim_{n\rightarrow\infty}\frac{1}{n}\sum_{j=1}^{n}\left(\ln\left|J_{1j}\right|-\ln\left|J_{2j}\right|\right)\right|
\end{eqnarray}
An analytical result can be obtained by taking the ensemble average
of the last expression. The limit of the sum turns into an integration
for independent and identically distributed disorder, 
\begin{equation}
\Lambda^{-1}=\frac{1}{4}\left|\int_{-1}^{1}d\omega\int_{-1}^{1}d\omega^{\prime}\left(\ln\left|J_{1}+\mu_{1}\omega\right|-\ln\left|J_{2}+\mu_{2}\omega^{\prime}\right|\right)\right|,\label{eq:loc_ssh}
\end{equation}
where $J_{1}$ and $J_{2}$ are the unperturbed hopping parameters.
$\mu_{1}$ and $\mu_{2}$ control the strength of disorder in $J_{1}$
and $J_{2}$ respectively. The random variables $\omega$ and $\omega^{\prime}$
are both drawn from a uniform distribution in the range $[-1,1]$,
leading to a normalization prefactor $1/4$. The analytic solution
to this integral has been obtained in \citep{Mondragon2014}, 
\begin{multline}
\Lambda^{-1}=\frac{1}{4\mu_{1}}\left[\left(J_{1}+\mu_{1}\right)\log\left|J_{1}+\mu_{1}\right|-\left(J_{1}-\mu_{1}\right)\log\left|J_{1}-\mu_{1}\right|\right]\\
-\frac{1}{4\mu_{2}}\left[\left(J_{2}+\mu_{2}\right)\log\left|J_{2}+\mu_{2}\right|-\left(J_{2}-\mu_{2}\right)\log\left|J_{2}-\mu_{2}\right|\right]\label{eq:2siteLL}
\end{multline}

%\begin{equation}%\Lambda^{-1}=\frac{\left(J_1+\mu _1\right) \log \left(\left\left| J_1+\mu _1\right\right| \right)-\left(J_1-\mu _1\right) \log \left(\left\left| J_1-\mu _1\right\right| \right)}{4\mu _1} %- \frac{\left(J_2+\mu _2\right) \log \left(\left\left| J_2+\mu _2\right\right| \right)-\left(J_2-\mu _2\right) \log \left(\left\left| J_2-\mu _2\right\right| \right)}{4\mu _2}. %\end{equation}

For small disorder, $\mu_{1},\mu_{2}\ll J_{2},J_{1}$, the localization
length Eq.~\eqref{eq:loc_ssh} can be approximated by 
\begin{eqnarray}
\Lambda^{-1} & \propto & \int_{-1}^{1}\int_{-1}^{1}d\omega_{1}d\omega_{2}\ln|J_{1}+\omega_{1}\mu_{1}|-\ln|J_{2}+\omega_{2}\mu_{2}|\nonumber \\
 & = & \int_{-1}^{1}\int_{-1}^{1}d\omega_{1}d\omega_{2}\ln|J_{1}|+\frac{\omega_{1}\mu_{1}}{J_{1}}-\frac{1}{2}\Big(\frac{\omega_{1}\mu_{1}}{J_{1}}\Big)^{2}-\Big[\ln|J_{2}|+\frac{\omega_{2}\mu_{2}}{J_{2}}-\frac{1}{2}\Big(\frac{\omega_{2}\mu_{2}}{J_{2}}\Big)^{2}\Big]\nonumber \\
 & + & \mathcal{O}\Big(\Big(\frac{\mu_{1}}{J_{1}}\Big)^{3}\Big)+\mathcal{O}\Big(\Big(\frac{\mu_{2}}{J_{2}}\Big)^{3}\Big).
\end{eqnarray}

Performing the integration up to order $\mathcal{O}\big(\big(\frac{\mu_{1}}{J_{1}}\big)^{3}\big)$
and $\mathcal{O}\big(\big(\frac{\mu_{2}}{J_{2}}\big)^{3}\big)$, one
finds that the localization length diverges for 
\begin{equation}
|J_{1}|(\mu_{1},\mu_{2})=|J_{2}|\exp\Big(\frac{J_{2}^{2}\mu_{1}^{2}-J_{1}^{2}\mu_{2}^{2}}{J_{1}^{2}J_{2}^{2}}\Big),
\end{equation}
which, up to leading order in the expansion of the exponential function,
reduces to 
\begin{equation}
|J_{1}|(\mu_{1},\mu_{2})=|J_{2}|\exp\Big(\frac{\mu_{1}^{2}-\mu_{2}^{2}}{J_{2}^{2}}\Big).
\end{equation}
We thus arrive at the conclusion that the value of $J_{1}$ where
the non-trivial$\leftrightarrow$trivial transition occurs increases
(decreases) compared to the clean case for small disorder strengths
if $\mu_{2}<\mu_{1}$ ($\mu_{2}>\mu_{1}$). This corresponds to the
\textit{topology by disorder} effect discussed in the main text and can be nicely seen
in Fig.~\ref{fig:pd_ssh}(b). For $\mu_{1}=\mu_{2}$, the phase transition
always occurs at $J_{1}=J_{2}$, as observed in Fig.~\ref{fig:pd_ssh}(a).
Now we continue to generalize the result to the non-Hermitian
trimer model. In Sec.~\ref{ds3site}, we have shown that the trimer model
of length $N\mod{3}$ can host two dark edge modes with energies $E=\pm J_{1}$.
These edge modes are supported purely by non-decaying sublattices.
The wave functions of the two dark states for disordered three-site
model is given by 
\begin{equation}
\psi_{n,ds\pm}=(-1)^{n-1}\prod_{j=1}^{n}\left|\frac{J_{2j}\pm J_{j}}{J_{3j}}\right|\psi_{1,ds\pm},
\end{equation}
where $J_{j}$, $J_{2j}$ and $J_{3j}$ are perturbed hopping parameters
in the $j$th unit cell. Since there exist two different edge modes,
we would expect two disjoint localization lengths, 
\begin{align}
\Lambda_{ds\pm}^{-1} & =-\lim_{n\rightarrow\infty}\frac{1}{n}\log\left|\psi_{n,ds\pm}\right|\\
 & =\left|\lim_{n\rightarrow\infty}\frac{1}{n}\sum_{j=1}^{n}\left(\ln\left|J_{2j}\pm J_{j}\right|-\ln\left|J_{3j}\right|\right)\right|.
\end{align}
Again, we take the ensemble average, and the summation turns into
an integration, which gives 
\begin{equation}
\Lambda_{ds\pm}^{-1}=\frac{1}{8}\left|\int_{-1}^{1}d\omega\int_{-1}^{1}d\omega^{\prime}\int_{-1}^{1}d\omega^{\prime\prime}\left(\ln\left|(J+\mu\omega)\pm(J_{2}+\mu_{2}\omega^{\prime})\right|-\ln\left|J_{3}+\mu_{3}\omega^{\prime\prime}\right|\right)\right|.
\end{equation}
Here $J$, $J_{2}$ and $J_{3}$ are unperturbed hopping parameters.
$\mu$ $\mu_{2}$ and $\mu_{3}$ define the amplitudes of disorder.
$\omega$, $\omega^{\prime}$ and $\omega^{\prime\prime}$ are three
independent and identically distributed random variables in the range
of $[-1,1]$. After performing the integration explicitly, we arrive
at 
\begin{eqnarray}
\Lambda_{ds\pm}^{-1} & = & {2\mu\mu_{2}}\Big\{\left(J\pm J_{2}-\mu-\mu_{2}\right)^{2}\log\left(\left|J\pm J_{2}-\mu-\mu_{2}\right|\right)-\left(J\pm J_{2}+\mu-\mu_{2}\right)^{2}\log\left(\left|J\pm J_{2}+\mu-\mu_{2}\right|\right)\nonumber \\
 & - & \left(J\pm J_{2}-\mu+\mu_{2}\right)^{2}\log\left(\left|J\pm J_{2}-\mu+\mu_{2}\right|\right)+\left(J\pm J_{2}+\mu+\mu_{2}\right)^{2}\log\left|J\pm J_{2}+\mu+\mu_{2}\right|\Big\}\nonumber \\
 & - & \frac{1}{\mu_{3}}\Big\{\left(J_{3}+\mu_{3}\right)\log\left(\left|J_{3}+\mu_{3}\right|\right)-\left(J_{3}-\mu_{3}\right)\log\left(\left|J_{3}-\mu_{3}\right|\right)\Big\}-4.\label{eq:3siteLL}
\end{eqnarray}
Eq. \eqref{eq:2siteLL} and Eq. \eqref{eq:3siteLL} allow us to trace
the exact critical phase transition contours in the non-Hermitian
SSH dimer and trimer models.

\section{Diagonal Disorder}
\label{sec:diagonal_disorder}
\begin{figure}
\begin{centering}
\includegraphics[width=0.4\textwidth]{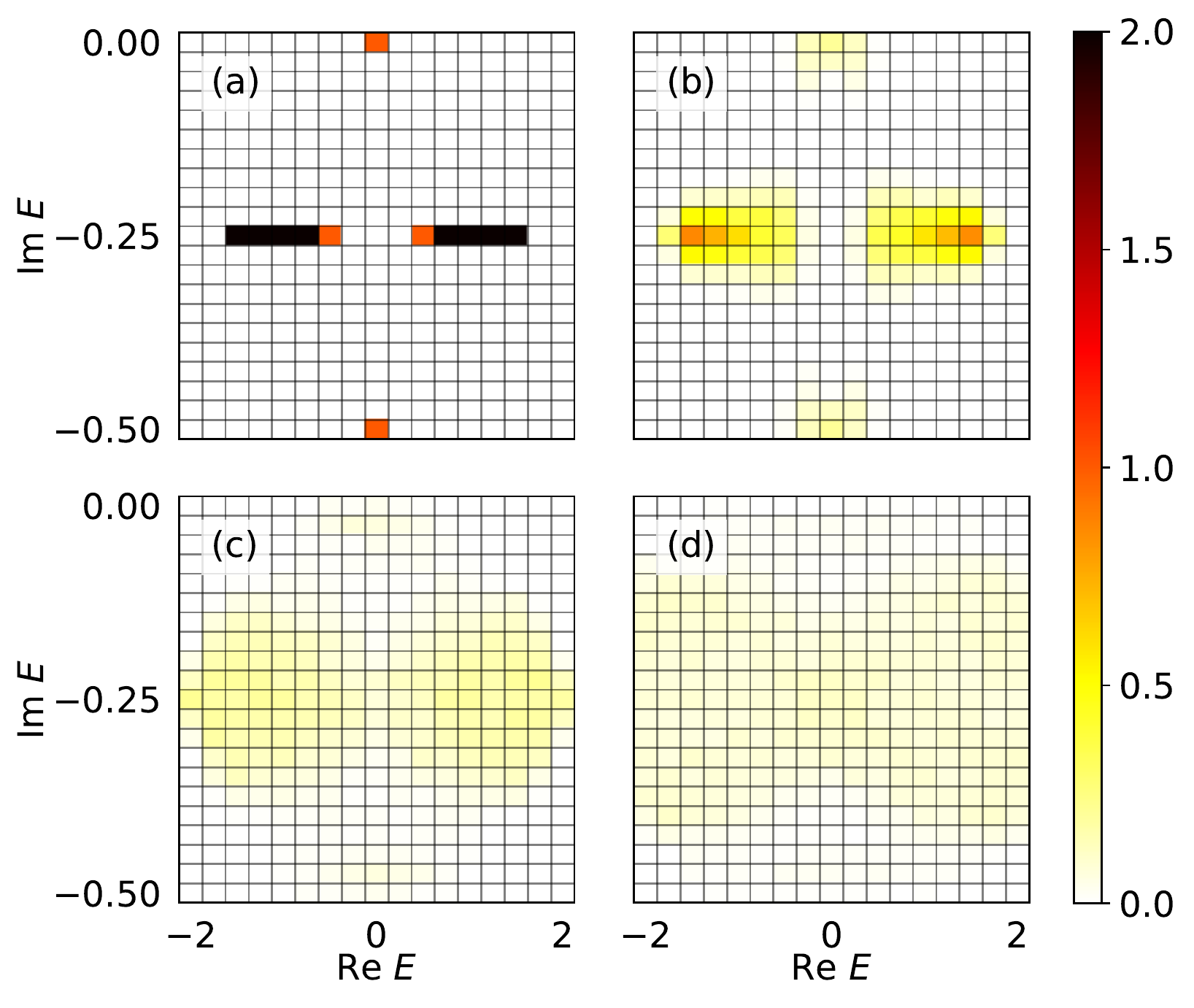}\includegraphics[width=0.4\textwidth]{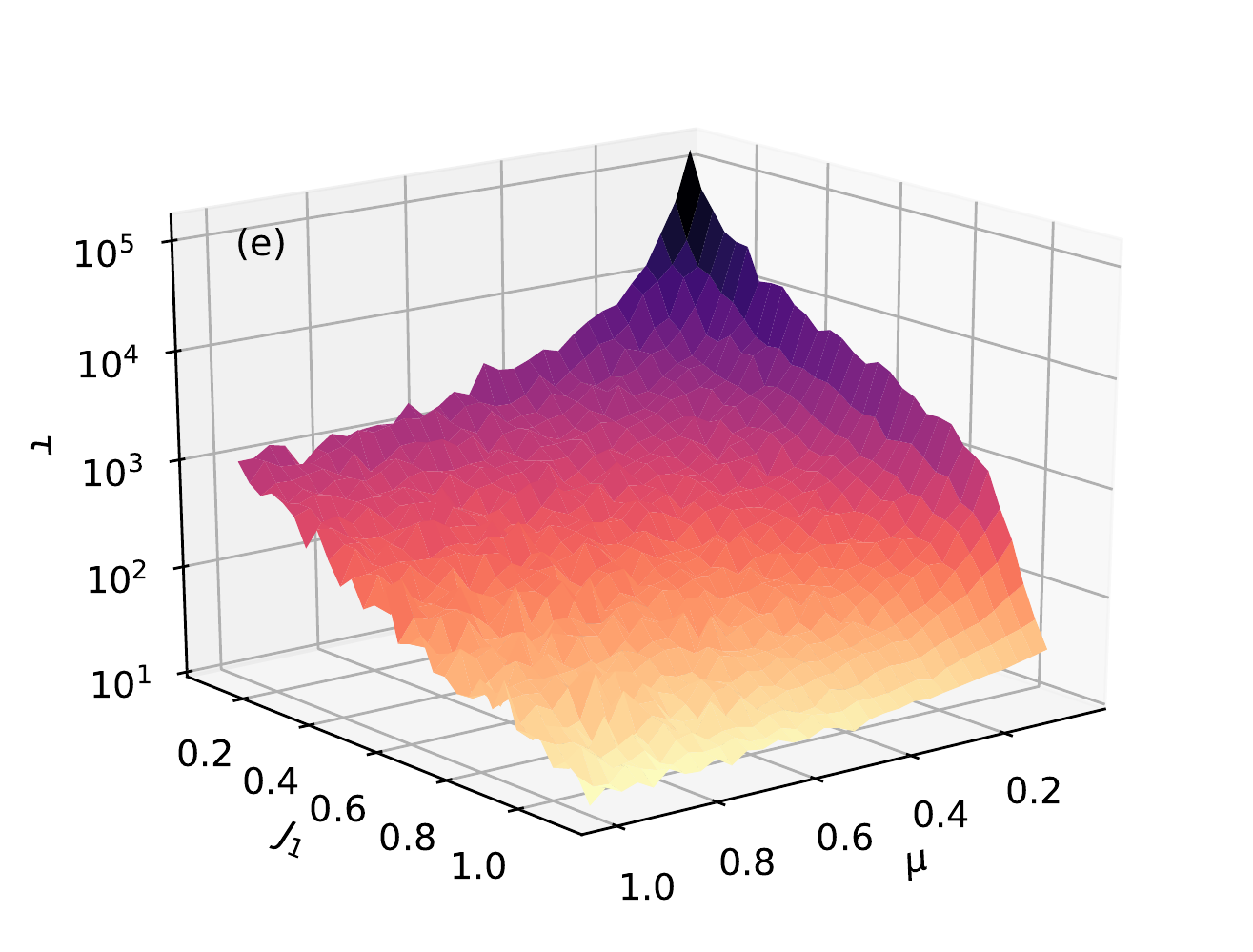}
\par\end{centering}
\caption{Eigenspectrum density of states of the restricted Hamiltonian $\mathsf{H}$
in the topologically non-trivial regime for (a) the clean system,
and diagonal disorder strengths (b) $\mu=0.5$, (c) $\mu=1$, (d)
$\mu=1.5$. Results are averaged over 1000 diagonalizations. Here,
$N=20$, $\Gamma=0.5$, $J_{1}=0.5$, $J_{2}=1$. (e) Coherence time
of the edge qubit as a function of the intra-cell hopping strength
$J_{1}$ and diagonal disorder $\mu$. In this setting, $J_{2}=1$,
$N=100$, and the time evolution of the coherence is disorder averaged
over 50 realizations. \label{fig:diag_disorder}}
\end{figure}

As argued in the main text, diagonal disorder destroys the protective
chiral symmetry of the SSH model, making it collapse to a topologically
trivial phase. This effect can be nicely seen when considering the
eigenspectrum density of states of the restricted Hamiltonian, as
already introduced in the main text for symmetry conserving disorder.
In analogy to off-diagonal disorder, the on-site potentials $\epsilon_{A,i}$
and $\epsilon_{B,i}$ are chosen to be uniformly distributed between
$[-\mu,\mu]$. %\begin{figure} %   \centering %  \includegraphics[width=0.4\textwidth]{SSH_dos_diag.pdf}
%  \caption{Eigenspectrum density of states of the restricted Hamiltonian $\mathsf{H}$ for (a) the clean system, and diagonal disorder strengths (b) $\mu=0.5$, (c) $\mu = 1.0$, (d) $\mu=1.5$. Results are averaged over 1000 diagonalizations. Here, $N=20$, $\Gamma=0.2$, $J_1=1$ and $J_2=1.4$.}
% \label{fig:dos_diag}%\end{figure}

% \begin{figure}%%    \centering%   \subfloat{{\includegraphics[width=0.4\textwidth]{SSH_dos_diag.pdf} }}%%    \qquad%    \subfloat{{\includegraphics[scale=0.6]{coh_time_diag_dis.pdf} }}%%    \caption{Eigenspectrum density of states of the restricted Hamiltonian $\mathsf{H}$ in the topologically non-trivial regime for (a) the clean system, and diagonal disorder strengths (b) $\mu=0.5$, (c) $\mu = 1.0$, (d) $\mu=1.5$. Results are averaged over 1000 diagonalizations. Here, $N=20$, $\Gamma=0.5$, $J_2=1$ and $J_1=0.5$. (e) Coherence time of the edge qubit as a function of the intra-cell hopping strength $J_1$ and diagonal disorder $\mu$. In this setting, $J_2=1$, $N=100$, and the time evolution of the coherence is disorder averaged over 50 realizations.}%%    \label{fig:diag_disorder}%%\end{figure}Figure
Fig.~\ref{fig:diag_disorder}~(a)-(d) illustrates how the topological dark
states appearing in the clean system quickly wash out, joining the
non-topological bulk state manifold. This is in in stark contrast
to a finite symmetry conserving off-diagonal disorder, where the topological
dark states were almost unaffected by the noise, cf. Figure \ref{fig:dos_offdiag}.

To underline the destructive effect further, the edgequbit's coherence
is inspected. As soon as disorder on the on-site potentials is introduced,
the coherence time is not infinite anymore, but it is reduced to a
finite value $\tau$. By assuming an exponential decay in time, i.e.,
$\mathcal{C}(t)=\mathcal{C}(t_{0})e^{-(t-t_{0})/\tau}$ for some $t_{0}\gg1/\Gamma$,
we can extract $\tau$ by integrating over the time evolution of the
coherence, i.e., 
\begin{equation}
I:=\int_{t_{0}}^{t_{1}}\mathcal{C}(t)dt=\int_{t_{0}}^{t_{1}}\mathcal{C}(t_{0})e^{-(t-t_{0})/\tau}dt=\tau(\mathcal{C}(t_{1})-\mathcal{C}(t_{0})).
\end{equation}

\begin{figure}[t]
\begin{centering}
 \includegraphics[scale=0.5]{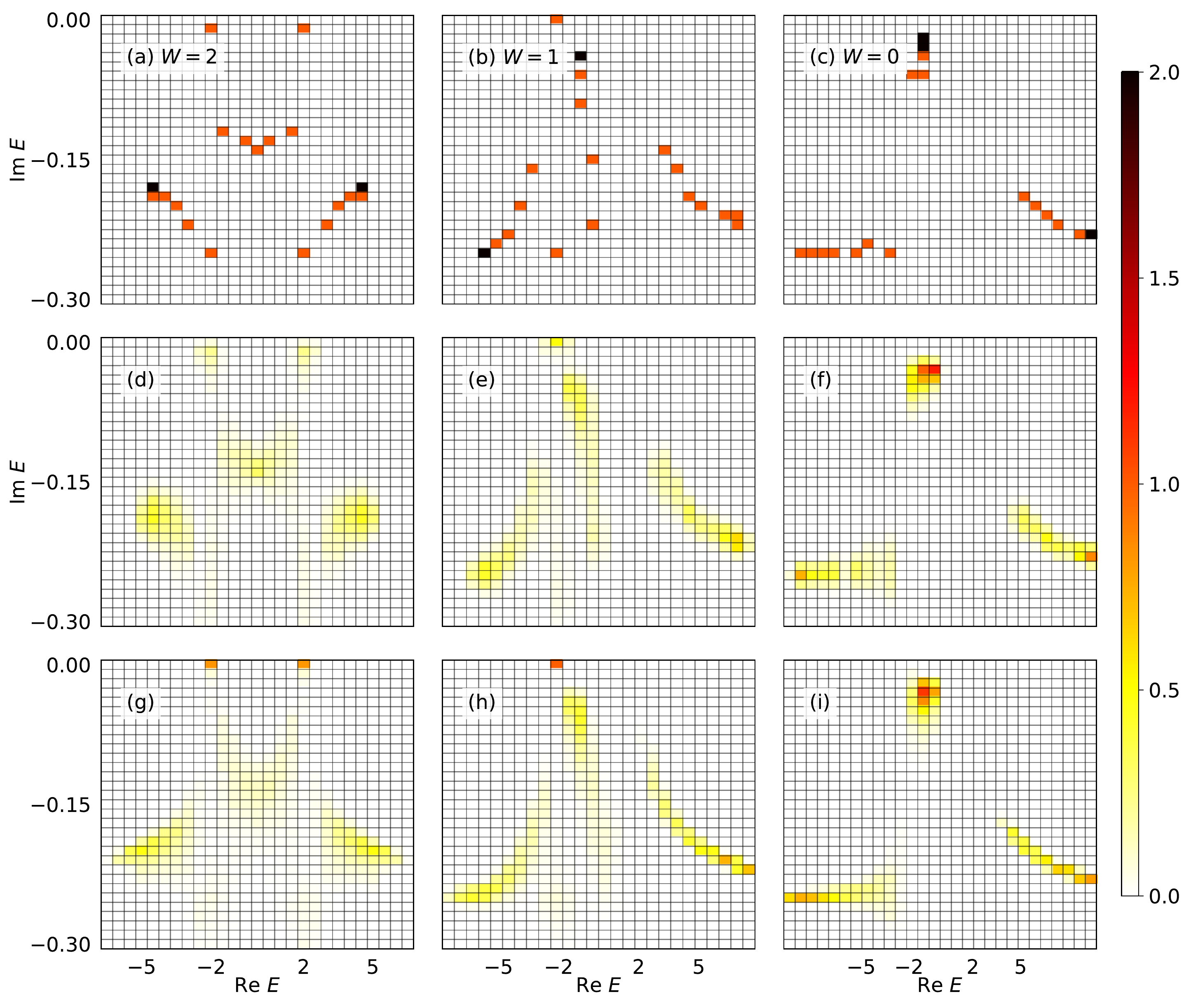} \caption{Density of states for the trimer model. (a)-(c) Clean density of states
for $W=2,1,0$, respectively. (d)-(f) Diagonal disorder $\mu=1$.
(g)-(i) Off diagonal disorder on $J_{2},J_{3},J$ with $\mu=1$. Here,
$N=21$, $J_{1}=J_{2}=2$, $J_{3}=3$, and $J=0,3,6$ for the topological
phases $W=2,1,0$, respectively. It is seen how $W$ quasi-dark states
with energies $E=\pm J_{1}$ exist in the clean system, being unstable
(stable) for the considered diagonal (off-diagonal) disorder. \label{fig:trimer_dos} }
\par\end{centering}
\end{figure}

Numerical integration leads to the results depicted in Fig.~\ref{fig:diag_disorder}~(e),
where a sharp drop of the coherence time away from the fully dimerized,
clean limit can be observed (notice the logarithmic scaling on the
z-axis). For the trimer model, very similar behavior is being observed,
for disorder acting on either on-site potentials or the coupling parameter
$J_{1}$, see Fig.~\ref{fig:trimer_dos} for the DOS.

\end{document}